\documentclass[preprint,12pt,3p]{elsarticle}

\usepackage{amssymb}
\usepackage{graphicx} 
\usepackage{array} 
\usepackage[table,xcdraw]{xcolor} 
\usepackage{multirow}
\usepackage{booktabs} 
\usepackage{amsmath}
\usepackage{amsfonts}
\usepackage{float}
\usepackage{xcolor}
\usepackage{colortbl}
\usepackage{multirow}
\usepackage{mathtools}
\usepackage{algorithm}
\usepackage{algorithmicx}
\usepackage{algpseudocode}
\usepackage{tabto}
\usepackage{todonotes}
\usepackage{hhline}
\usepackage{enumitem}
\usepackage{caption}
\usepackage{subfigure}

\makeatletter
\def\ps@pprintTitle{%
 \def\@oddhead{\leftline{\textit{Accepted at Elsevier Journal of Microprocessors and Microsystems (MICPRO)}}}%
 \let\@evenhead\@empty
 \def\@oddfoot{\centerline{\thepage}}%
 \let\@evenfoot\@oddfoot}
\makeatother

\usepackage{tikz}

\biboptions{sort&compress}
\makeatletter

\begin{document}
\let\today\relax
\begin{frontmatter}

\title{\vspace*{2pt}SIMCom: Statistical \underline{S}niffing of \underline{I}nter-\underline{M}odule \underline{Com}munications for Runtime Hardware Trojan Detection}

\author[label1]{Faiq Khalid\corref{cor1}}
\ead{faiq.khalid@tuwien.ac.at}
\author[label2]{Syed Rafay Hasan}
\ead{shasan@tntech.edu}

\author[label3]{Osman Hasan}
\ead{osman.hasan@seecs.edu.pk}

\author[label1]{Muhammad Shafique}
\ead{muhammad.shafique@tuwien.ac.at}

\address[label1]{Technische Universit\"at Wien (TU Wien), Vienna, Austria}
\address[label2]{Department of Electrical and Computer Engineering, Tennessee Tech University, Cookeville, TN, USA}
\address[label3]{National University of Sciences and Technology (NUST), Islamabad, Pakistan}

\cortext[cor1]{Corresponding author}




\begin{abstract}
	Timely detection of Hardware Trojans (HTs) has become a major challenge for secure integrated circuits. We present a run-time methodology for HT detection that employs a multi-parameter statistical traffic modeling of the communication channel in a given System-on-Chip (SoC), named as SIMCom. The main idea is to model the communication using multiple side-channel information like the Hurst exponent, the standard deviation of the injection distribution and the hop distribution jointly to accurately identify HT-based online anomalies (that affects the communication without affecting the protocols or control signals). At design time, our methodology employs a ``property specification language'' to define and embed assertions in the RTL, specifying the correct communication behavior of a given SoC. At run-time, it monitors the anomalies in the communication behavior by checking the execution patterns against these assertions. For illustration, we evaluate SIMCom for three SoCs, i.e., SoC1 ( four single-core MC8051 and UART modules), SoC2 (four single-core MC8051, AES, ethernet, memctrl, BasicRSA, RS232 modules), and SoC3 (four single-core LEON3 connected with each other and AES, ethernet, memctrl, BasicRSA, RS23s modules microcontrollers). The experimental results show that with the combined analysis of multiple statistical parameters, SIMCom is able to detect all the benchmark Trojans (available on trust-hub) with less than 1\% area and power overhead.  
\end{abstract}


\end{frontmatter}

\section{Introduction}\label{introduction}
Globalization encourages the use of the intellectual property (IP)-based system-on-chip (SoC) designs and embedded microcontrollers. Especially, their applicability as the low-edge edge computing devices in the Internet-of-Things (IoT) scenarios, and low-end embedded computing nodes in the cyber-physical systems (CPS, like the industrial control systems), is widespread \cite{mohan2013s3a,levshun2019design,wang2018model,giakoumis2018chaos,ratasich2019roadmap,shafique2018intelligent,moura2019cyber}. However, this trend increases the chances of malicious hardware design intrusions, known as Hardware Trojans (HTs), into the modern-day SoCs \cite{tehranipoor2010survey,subramani2018hardware}. HTs can lead to several unwanted payloads, i.e., information leakage, change in the timing characteristics, malfunctioning and denial-of-service (DoS)~\cite{tehranipoor2010survey,subramani2018hardware}. The effects can be catastrophic, such as system failure and leakage of secret encryption keys (e.g., failure of an ice-detection module in the P-8A Poseidon~\cite{villasenor2013hidden} and chip insertion attack in 2018~\cite{bloomberg_security}, making it imperative to develop effective HT detection techniques.\par

\begin{figure}[!t]
	\centering
	\includegraphics[width=1\linewidth]{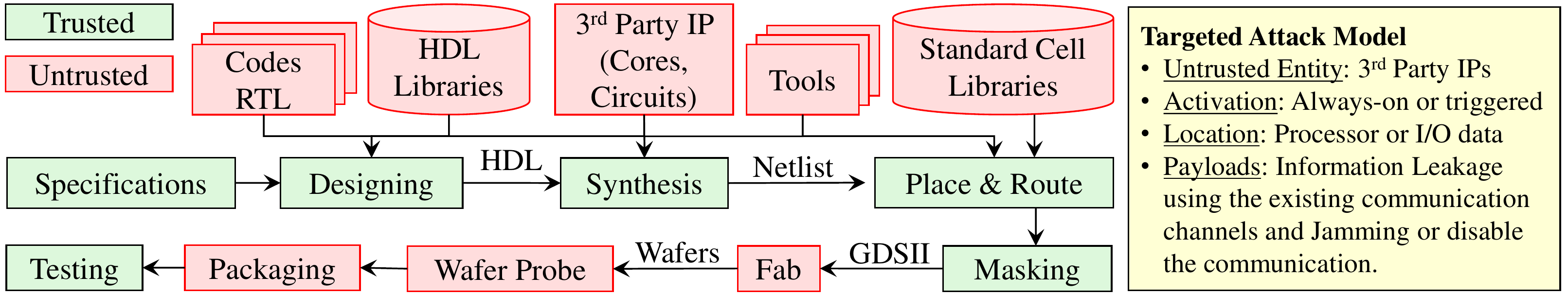} 
	\caption{Left side, all possible security vulnerabilities in IC Supply Chain. Right side, targeted attack model.}
	\label{fig:threat_model}
\end{figure}

Some of the contemporary HT detection techniques utilize the circuit timing behavior \cite{zarrinchian2017latch,plusquellic2018detecting,nandhini2018delay,fang2018prefetch,xue2018hardware}, power consumption  \cite{lodhi2017power,hoque2017golden,lodhi2016self,zhang2018data}, current or electromagnetic signals based \textbf{golden signatures} to detect anomalous electrical behavior \cite{gbade2014signature,lodhi2014hardware}. However, \textit{in case of the third-party-IP based designs, it is nearly impossible to extract the golden signatures}. Moreover, in most of the 3PIP-based SoCs, the facility to integrate the IPs can be trusted (as, shown in Figure~\ref{fig:threat_model}). On the other hand, it is difficult to \textit{guarantee} that the IP vendors can be trusted. To address this issue, various IP analysis-based approaches have been proposed to get the golden behavior \cite{zhang2014detrust,waksman2013fanci,zhang2015veritrust,haider2014hatch,ngo2015hardware,zareen2018detecting,cui2018hardware,liu2019hardware,cui2018hardware} but these techniques inherently pose the following limitations:\par
\begin{enumerate}
	\item \textit{Accuracy of estimated golden behavior:} Due to limited access to IPs, they use different estimation algorithms to extract the golden model \cite{he2017hardware}. However, due to measurement inaccuracies and behavioral estimation, they cannot guarantee the accuracy of the golden model. 
	\item  In the case of reverse engineering, sensors for golden data extraction \textit{cannot encompass all the possible input conditions} for larger ICs because of the inherent data loss during the quantization in analog-to-digital conversion \cite{pelgrom2017nyquist}.    
\end{enumerate} 

These limitations raise a key research question: \textit{How can signature-based HT detection techniques effectively work if an intruder (at foundry) exploits the intricacies of estimation algorithms to insert Trojans that remain dormant during the testing stages \cite{bhunia2014hardware}?} For example, in the SoC design, some hard or firm IPs may hide Trojans depending on the aging of the chip \cite{becker2013stealthy}, and they only get activated during run-time \cite{hasan2015tenacious,mossa2017hardware,mossa2017self}.\par

To address this issue, several run-time detection approaches have been developed to monitor a SoC for its entire operational lifetime, providing an important last-line of defense \cite{forte2013temperature,bao2015temperature,zhao2015applying,lodhi2017power,bao2016reverse,lodhi2016self,iwase2015detection}. However, most of these techniques are based on side-channel analysis (SCA), which require \textbf{precise calibration} for differentiating between the process variations and the intrusion behavior. They also rely on the premise that triggering of payload results in a substantially higher current flow. To avoid the complex requirements of the SCA-based run-time HT detection techniques, alternative behavior can be used to sniff the abnormalities during run-time. One of the most prominent ones is \textit{the communication behavior} because, in real-world scenarios, modules are connected via communication channels and therefore, most of the intrusions have an impact on the communication behavior without affecting the communication protocols. 

\subsection{Motivational Analysis}
\begin{figure}[!t]
	\centering
	\includegraphics[width=1\linewidth]{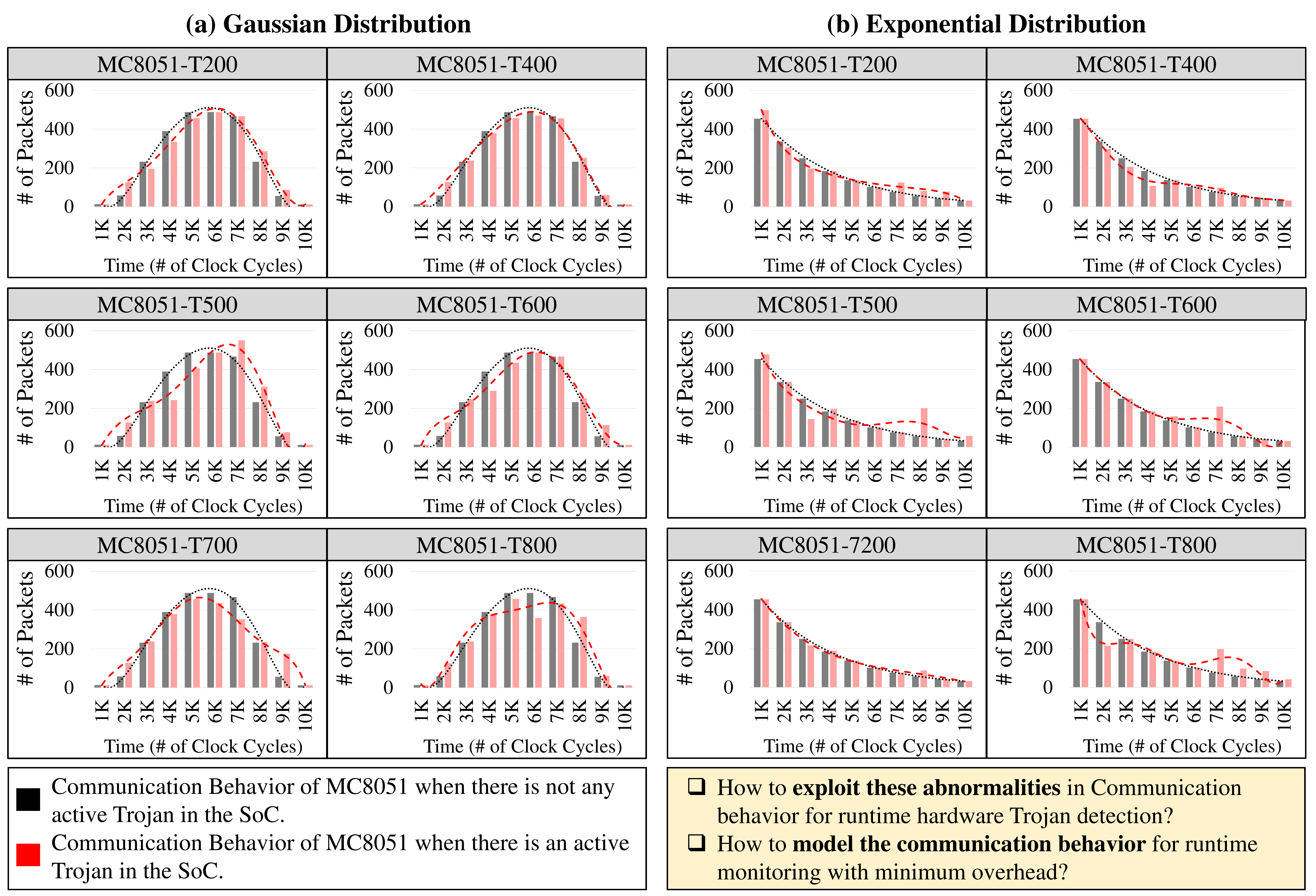} 
	\caption{Experimental analysis to study the effects of Trust-Hub HT benchmarks (i.e., MC8051-T200, T300, T400, T500, T600, T700, and T800) on the communication behavior of MC8051 in an SoC (it consists of four single-core MC8051 IPs that are linked with each other using AMBA 2.0 bus). Note, all these experiments are performed using two different input data distributions, i.e., Gaussian and Exponential distributions. This analysis shows that all the implemented HT benchmarks have a detectable impact on communication patterns of MC8051, which can be observed by analyzing the change in the shape of the communication distribution in MC8051. }
	\label{fig:com_behavior}
\end{figure}

To validate the aforementioned observation, we analyze the communication behavior of an MC8051 microcontroller for the Gaussian and exponentially distributed input data, in the presence of multiple trust-hub Trojan benchmarks, i.e., MC8051-T200, T300, T400, T500, T600, T700, and T800 \cite{trust-HUB}. Our analysis in Figure~\ref{fig:com_behavior} shows that all the benchmarks have some effects on the communication patterns. For example, MC8051-T600 slightly changes the communication behavior in case of the exponential distribution. However, in the case of Gaussian distribution, it significantly changes the communication behavior of the microcontroller. In short, this analysis shows that \textit{communication behavior can be monitored to identify abnormalities during run-time without estimating the functional behavior of a particular module}.

This observation leads to a research challenge about obtaining the golden statistical model in the presence of untrusted IPs. To address this challenge, either we assume that Trojans remain dormant during the design stage, fabrication stage, and testing stage, or we assume that at least on the IPs is trusted. However, in real-world applications, Trojans may be active during the design stage, fabrication stage, or testing stage. Therefore, in this paper, we assume that at least one of the IPs is trusted, and the defender deploys the defense inside the trusted IP or in the IP integration platform. Based on this assumption, we postulate a hypothesis, \textit{``information leakage using such communication channels, which are linked with a trusted IP can have a detectable impact on the statistical behavior of the trusted IP communication''.}

\begin{figure}[!t]
    	\centering
    	\includegraphics[width=1\linewidth]{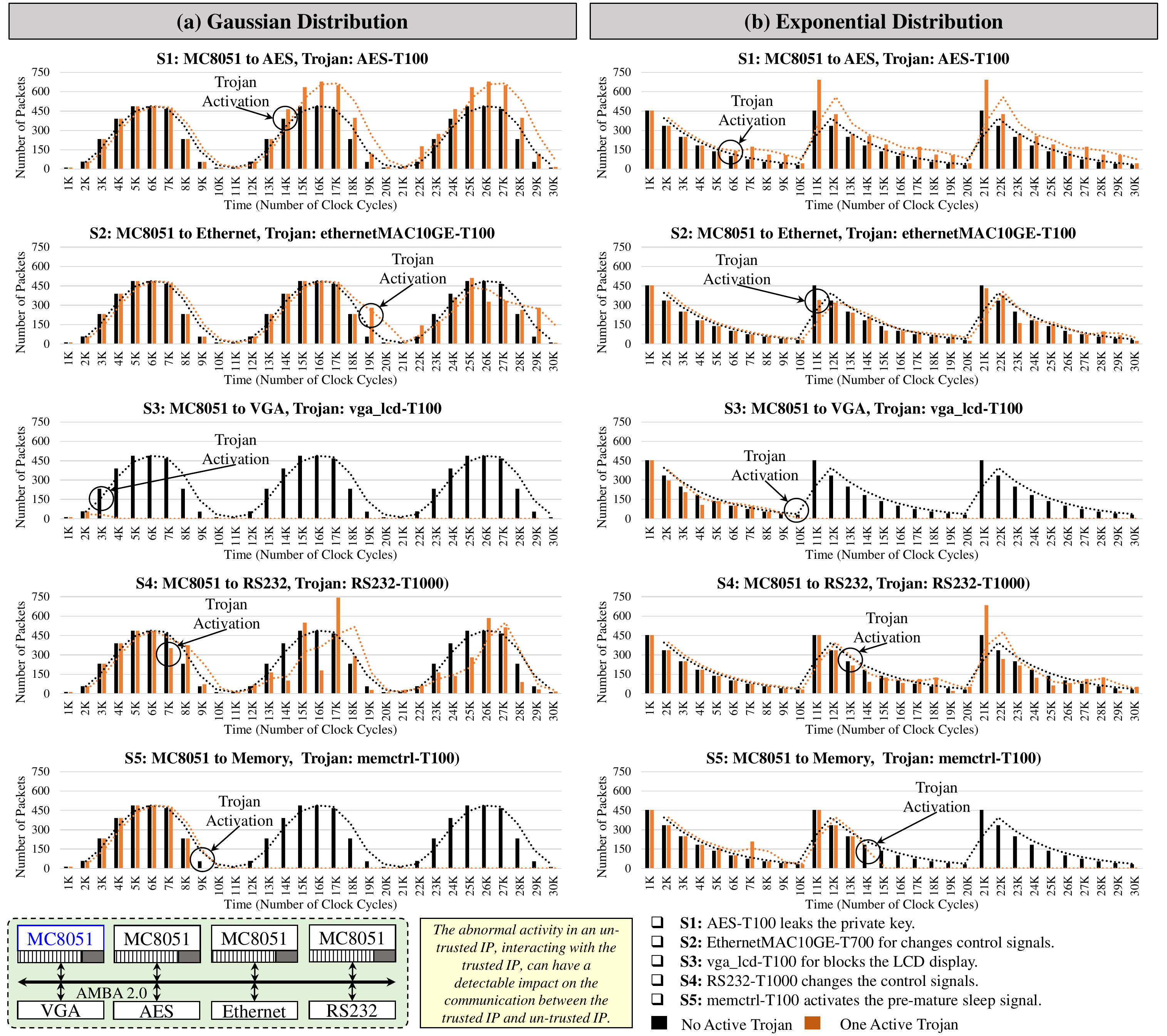} 
    	\caption{Experimental analysis to study the untrusted IPs on the communication of the trusted IPs. The SoC used in this analysis consists of four single-core MC8051 IPs that are linked with each other, and other IPs like RS232, vga\_lcd, ethernet, using AMBA 2.0 bus. Note, all these experiments are performed using two different input data distributions, i.e., Gaussian and Exponential distributions, and different Trust-Hub HT benchmarks (e.g., AES-T100, ethernetMAC10GE-T100, vga\_lcd-T100, RS232-T1000, memctrl-T100) with different payloads like information leakage and blocking/jamming the communication. In this analysis, only one MC8051 IP is trusted, and one HT benchmark is active at a time. This analysis shows that all the implemented HT benchmarks have a detectable impact on communication patterns of the trusted MC8051 IP, which can be observed by analyzing the change in the shape of the communication distribution of the trusted MC8051.}
    	\label{fig:com_behavior_2}
    \end{figure}

To illustrate this hypothesis, we perform a motivational case study about four single-core MC8051 interacting with multiple IPs, i.e., vga\_lcd, AES, Ethernet, RS232, memory controller, in an SoC, as shown in Figure~\ref{fig:com_behavior_2}. In this case study, we assume that only one single-core MC8051 is trusted, and the rest of the IPs are untrusted. For a comprehensive analysis, we implemented the multiple Trojans, i.e., AES-T100, ethernetMAC10GE-T100, vga\_lcd-T100, RS232-T1000, memctrl-T100, with different payloads like information leakage and blocking/jamming the communication. To study the minimum impact of the payload of the Trojan, we also assume that only one Trojan is active at a time. By analyzing the results in Figure~\ref{fig:com_behavior_2}, we made the following observations:

\begin{itemize}
        \item Communication jamming/blocking caused by un-trusted IP has a significant impact on the statistical properties of the trusted IP communication. For example, in the case of S3 and S5 of Figure~\ref{fig:com_behavior_2}, when Trojan gets the trigger in vga\_lcd or memctrl, it initiates the blank display (in the case of vga\_lcd-T100) or enables the premature sleep signal (in the case of memctrl-T100). Hence, it blocks communication with trusted MC8051.
        
        \item Constant information leakage through untrusted IP has a significant impact on the communication traffic, which eventually changes the statistical properties of the communication of the trusted IP. For example, in the case of S1 of Figure~\ref{fig:com_behavior_2}, the communication traffic increases after the Trojan (AES-T100) gets the trigger. Hence, it changes the statistical properties of the communication of the trusted MC8051. 
        
        \item If the control signals or internal protocol signals are compromised, then it also has a relatively less impact on communication with trusted IP. For example, in the case of S2 and S4 of Figure~\ref{fig:com_behavior_2}, the triggered Trojans sometimes increase the communication traffic and sometimes decreases the communication traffic depending upon the payload. Hence, it can also change some of the statistical properties of the communication of the trusted MC8051.

    \end{itemize}
The experimental analysis in Fig.~\ref{fig:com_behavior} shows that even the simple communication between MC8051 and UART maintains a distinct communication pattern. Therefore, we can safely conclude that it is possible to extract distinct communication patterns in both the high-end and low-end microcontrollers.  These communication patterns can be used to define an appropriate statistical model that represents the communication activities in both low-end and high-end microprocessors.
\subsection{Associated Research Challenges:} 
Based on these observations, we conclude that run-time monitoring of the communication behavior of the trusted IPs can be used to detect the abnormal activities of un-trusted IPs. This observation is only valid if the untrusted IPs are linked with trusted IP via active communication channels.  However, run-time monitoring of the communication behavior of the trusted IPs poses the following research challenges:

\begin{enumerate}
	\item \textbf{Modeling of communication behavior:} Unlike the side channel parameters, typical mathematical modeling techniques cannot be used to model communication behavior because of its probabilistic nature. One of the solutions is to use sophisticated statistical modeling of communication behavior, but it comes with huge overhead. Therefore, the main questions that need to be addressed are:
	\begin{itemize}
	    \item How to statistically model the communication behavior of the trusted IP that can be used for run-time monitoring with minimum overhead?
	    \item What is the minimum number of observations required to learn/extract a well-defined communication pattern?
        \item How to translate the extracted communication patterns to the respective statistical model?
        \item How to store and use the golden statistical model for run-time analysis?
	\end{itemize}
	
	\item \textbf{Extracting the golden communication behavior:} In run-time monitoring, the fundamental challenge is to extract the golden behavior during the design time that can effectively be used for HT detection. 
\end{enumerate}

\subsection{Our Novel Contributions}
\begin{figure}[!t]
    \centering
    \includegraphics[width=1\linewidth]{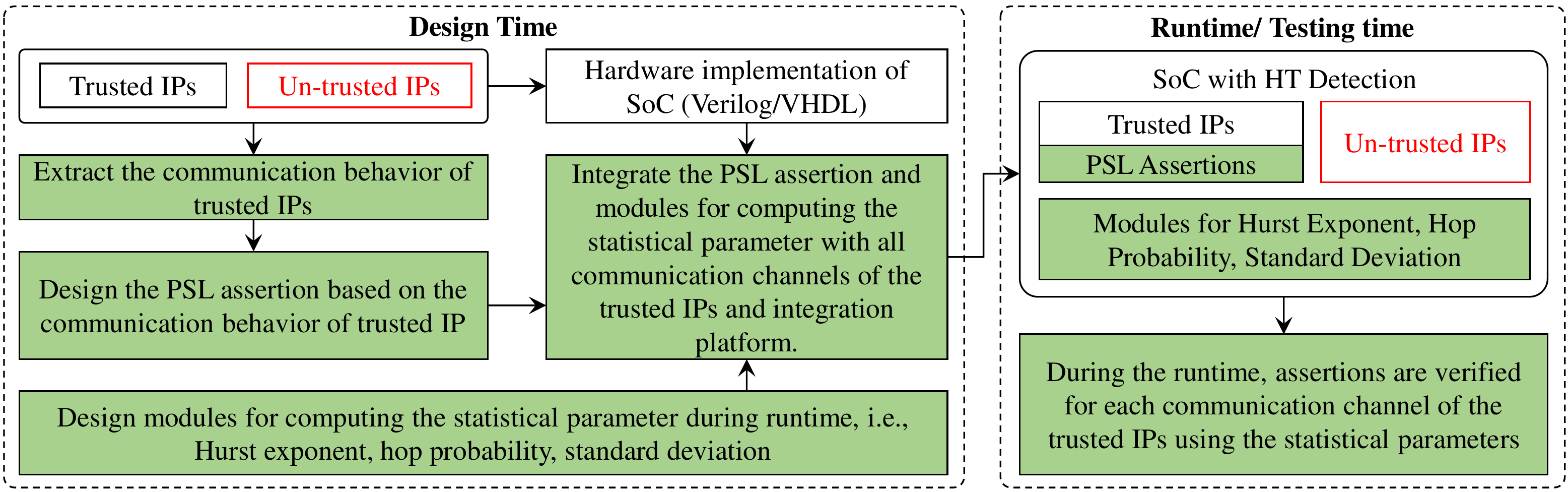} 
    \caption{Design, test and run-time flow of our methodology for statistical modeling of communication of the trusted IPs (SIMCom). The highlighted box represents novel contributions. The evaluation/testing is done on an MC8051-based SoC and a LEON3-based SoC.}
	\label{fig:Novel}
\end{figure}

To address above mentioned challenges, we propose a novel methodology that leverages the statistical traffic modeling of communication channels (\textit{SIMCom}, Section \ref{proposed-methodology}) in the SoC to sniff the possible anomalies in 3PIP units \footnote{The archive version of original submission is also available online: F. Khalid, et al. "SIMCom: Statistical Sniffing of Inter-Module Communications for Run-time Hardware Trojan Detection." arXiv preprint arXiv:1901.07299 (2018).}. The proposed methodology consists of the following three key steps, shown in Figure~\ref{fig:Novel}. \par
\begin{enumerate}
    \item \textbf{Statistical modeling of communication behavior (Section \ref{TM-8051_model}):} To extract the golden communication behavior that can be used during the run-time for HT detection. We assume that in an SoC at least one of the IPs is trusted, and by analyzing the communication of the trusted IP, we can extract the requires golden behavior. Therefore, We propose to statistically model the normal communication traffic of the trusted IP, during the design phase of the SoC. This behavior is extracted by obtaining a statistical traffic model, which is characterized based on the following observations:  
	\begin{enumerate}
	\item How often the packets are injected in the communication channel? 
	\item On average, how far does each packet travel?
	\item What portion of the total traffic has been injected in the communication channel by each module?
	\end{enumerate} 

	\item To identify the appropriate statistical model, we propose to choose the Hurst exponent because of the following reasons:
    \begin{itemize}
        \item It is very sensitive to the distribution of the time series. 
        \item It requires very few observations to estimate the Hurst exponent. For example, the minimum observations required for Hurst exponent is approximately 240. 
    \end{itemize} 
	\item \textbf{Run-time monitoring:} To design the run-time monitoring setup, we propose to translate the statistical model of communication behavior into their corresponding property specification language (PSL) assertions (see Sections \ref{PSL} and \ref{PSL_run} ). These assertions are embedded into the RTL description of SoC along with the traffic monitoring units. During the run-time or testing time, these assertions are verified based on the run-time values of the statistical parameters, i.e., Hurst exponent, hop probability, standard deviation.
\end{enumerate}

\normalcolor
To evaluate the effectiveness of SIMCom, we implemented three SoCs. First, SoC consists of four single-core MC8051 that are linked with universal asynchronous receiver/transmitter (UART) modules via and AMBA 2.0 bus. Second SoC consists of four single-core MC8051 IPs that are linked with each other and other IPs like RS232, vga\_lcd, ethernet, using AMBA 2.0 bus. Third SoC consists of four single-core LEON3 IPs that are linked with each other and other IPs like RS232, vga\_lcd, ethernet, using AMBA 2.0 bus. The experimental results show that with the help of multiple parameters, SIMCom is able to detect all the implemented Trojan benchmarks \cite{trust-HUB} with less than 1\% area and power overhead. 

\subsection{Paper organization}
The rest of the paper is organized as follows: Section~\ref{pre_liminaries} provides the provide a brief overview of statistical modeling of communication behavior using Hurst exponent, the standard deviation of the injection distribution and the hop distribution. Section~\ref{proposed-methodology} elaborates on the proposed methodology based on the statistical traffic modeling of SoC communication network for run-time hardware Trojan detection. Section~\ref{sec:HW} provides the details about implementations of the hardware modules that compute statistical parameters during run time. For illustration, Section~\ref{case-study} present the detailed analysis of three SoCs, i.e., SoC1 consisting of four single-core MC8051 with UART modules, SoC2 consisting of four single-core MC8051 linked with each other and AES, ethernet, memctrl, BasicRSA, RS23s modules, and SoC3 consisting of four single-core LEON3 connected with each other and AES, ethernet, memctrl, BasicRSA, RS23s modules. Section \ref{comparison} discusses the detection approaches, attributes, pros and cons, and a brief comparison w.r.t HT detection accuracy of the proposed approach and the state-of-the-art techniques. it  Finally, Section \ref{conclusion} concludes the paper. 	

\section{Preliminaries}
\label{pre_liminaries}
Before proceeding to the details of the technical contributions, in this section, we provide a brief overview statistical modeling of communication behavior.
\subsection{Statistical Traffic Modeling}\label{Statistical-Traffic-Modeling}
In order to extract the statistical communication behavior of the SoC, which can be used in turn to sniff any anomaly, we utilized the statistical model for SoC traffic provided in \cite{soteriou2006statistical}. The model is composed of the following three key statistical parameters: \par
\begin{enumerate}[leftmargin=*]
	\item \textit{\textbf{Hurst Exponent ($H$):}} Also known as the ``index of dependence'' or `` the index of long-range dependence'', it quantifies the relative tendency of a time series either to regress strongly to the mean or to cluster in a direction \cite{kleinow2002testing}. Typically H value ranges from 0 to 1. A value from 0.5 to 1 indicates that a high value in the series will probably be followed by another high value. A value from 0 to 0.5 indicates that it is more likely that a high value will be followed by a low value. A value of H equal to 0.5 indicates a completely uncorrelated series. The Hurst exponent can be estimated by using the following equation:
	\begin{equation}
	H = \dfrac{\log_{10}{\left( E\left[\dfrac{R(s)}{S(n)}\right] \right) }}{a_2 \times \log_{10} n}
	\label {eq:eq1}
	\end{equation}
	Where:\par
	$R(s)$: Magnitude range of the time series\par
	$S(n)$: Average peak magnitude of the time series\par
	$n$: Total number of observations\par
	$a_2$: Positive constant\par 
	In the traffic modeling of an SoC, the ``\textit{H}'' exponent corresponds to the magnitude of the packet injection peaks and their time arrival injection patterns.
	\item \textit{\textbf{Spatial Hop Distribution ($P$):}} It is determined by three factors: (1) the communication pattern of an application, (2) the power consumption of each communicating module and (3) the number of communicating modules required to complete an application. It is possible that the communication pattern and power consumption are fixed for a particular application, but the number of communicating modules, for an application, can vary and affect the hop counts. Therefore, Soteriou et. al. \cite{soteriou2006statistical} proposed a parameter, called \textit{acceptance probability ($p$)} of the packet, over an average hop count as a metric for equivalent spatial traffic distribution. For a total $N$ number of modules, if an application is mapped on $n$ modules, with $n \leq N$, then the acceptance probability is defined as $p = \dfrac{1}{n-1}$. In reality, a packet may only have a limited number of receivers $r$,  $n \geq N$, then the acceptance probability is $p = \dfrac{1}{r-1}$. The hop distance also depends upon the communication topology and thus the spatial hop distribution ($P_{h>d}$) for a particular $p$ can be defined as:\par
	\begin{equation}
	P_{h>d} = (1- p)^{s(d)}
	\label {eq:eq2}
	\end{equation}
	Where, if $s$ represents a source module, then $n(d)$, $s(d)$ and $d$ represent, the total number of modules whose hop distance (Manhattan) is $d$ from $s$, number of modules with hop distance $1$ to $d$, and the maximum hop distance, respectively. The values for $n(d)$ and $s(d)$ for different communication topologies are given in Table \ref{tab:tab1}.
	\item \textit{\textbf{Probability Distribution of Spatial Injection ($\sigma$):}} It represents the standard deviation of the probability distribution at which a module injects the packets into the communication network of an SoC. In this work, we used the Gaussian probability distribution.        
\end{enumerate}
	\begin{table}[H]
	\centering
	\caption{$n(d)$ and $s(d)$ for different SoC communication topologies \cite{soteriou2006statistical}}
	\label{tab:tab1}
	\resizebox{0.7\textwidth}{!}{
		\begin{tabular}{|c|c|c|}
			\hline
			\textbf{Topology} & \textbf{$n(d)$} & \textbf{$s(d)$}  \\ \hline
			Ring & $2$ & $2d$ \\ \hline
			2D Mesh & $4d$ & $2d \times (d+1)$  \\ \hline
			3D Mesh & $(8\sum_{a=1}^{d} (d-a)) + 4d + 2$ & $\sum_{a=1}^{d} 4a \times (2d-1a+1) + 2d$  \\ \hline
	\end{tabular}}
	\vspace{0.1in}
\end{table} 
\section{SIMCom: Statistical Sniffing of Communication Behavior} \label{proposed-methodology}

This section explains our proposed methodology for runtime hardware Trojan detection based on the traffic modeling of SoC communication.
Using an appropriate threat model is one of the foremost steps in developing any methodology for detecting intrusions in the domain of hardware security. In this paper, we assume that the 3PIP vendors are not trusted, and the SoC integration is performed in a trusted facility \cite{xiao2016hardware}. Moreover, we also assume that at least one 3PIP has to be trusted, which can be built in-house, such that its VHDL is available to insert checks\footnote{Note, If none of the 3PIPs are trusted, then finding Trojan is an extremely hard problem as no hardware access is known, and no golden behavior is known.}. 

\begin{figure*}[!t]
	\centering
	\includegraphics[width=1\textwidth]{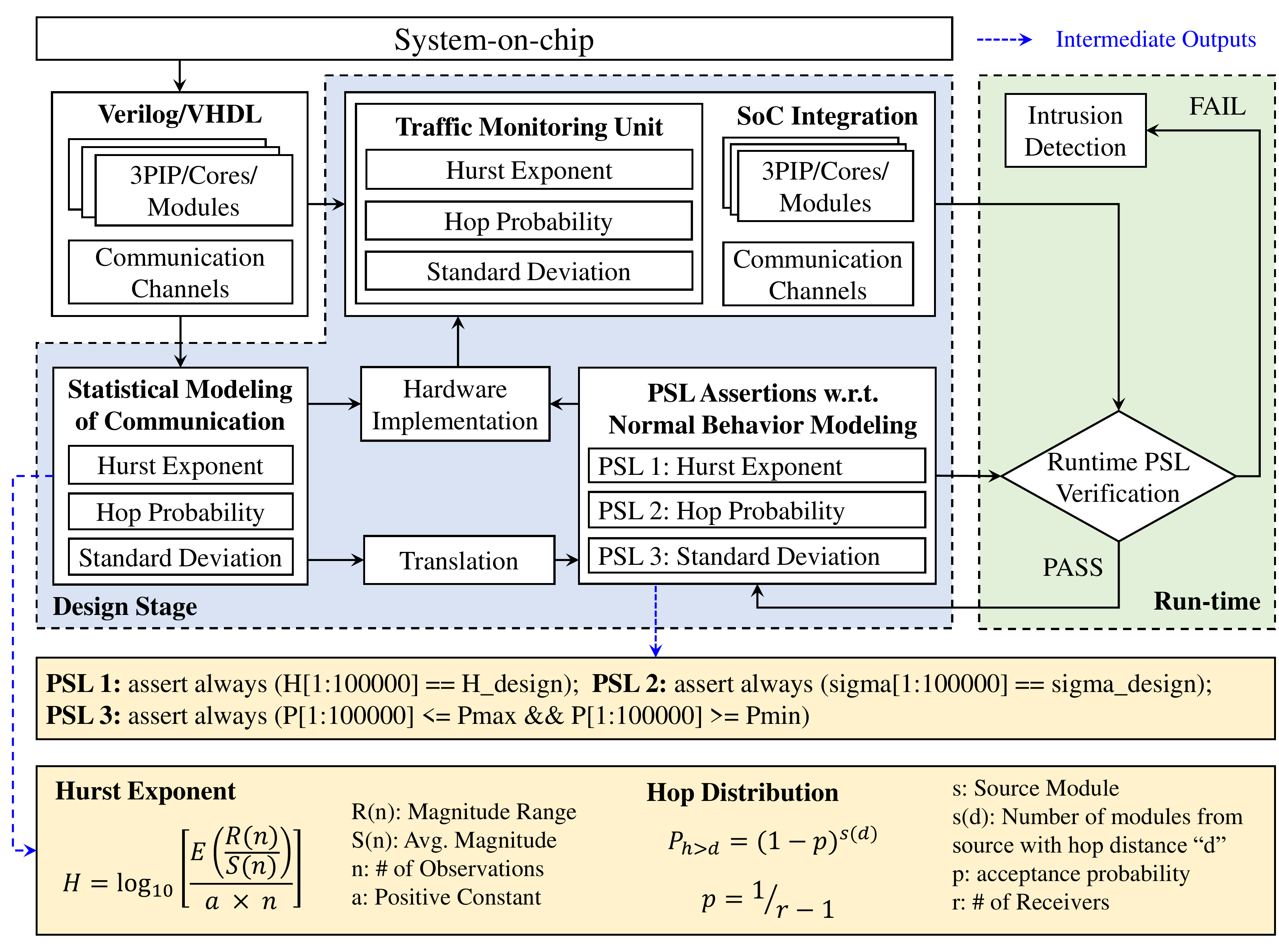}
	\caption{SIMCom: Statistical sniffing of inter-module communication for runtime HT detection}
	\label{fig:methodology}
\end{figure*}

SIMCom consists of two major phases, as depicted by the dotted rectangles in Figure~\ref{fig:methodology}. 

\textbf{During the design time}, SIMCom requires the following steps to design the runtime monitors:
\begin{enumerate}
    \item First, it extracts the communication behavior of the trusted IP and statistically models the extracted communication behavior of the trusted IP. Note, the traffic modeling is done under the premise that at least one of the IP is trusted. The statistical modeling of the trusted IP can be used to detect HTs. The statistical communication behavior of the trusted IP in an SoC is obtained using the following statistical parameters.

    \begin{enumerate}
        	\item The input packets are generated with respect to the standard spatial injection distribution, e.g., Gaussian distribution. 
        	\item Next, the \textit{Hurst exponent} is computed for each communication channel using the R/S method. 
        	\item Hop distribution is computed using the total number of available communication channels and an active communication channel.  
        	
        \end{enumerate}
    
    \item Then, it uses the statistical model to define the corresponding property specification language assertions. These assertions are inserted into the RTL or Verilog/VHDL of the trusted IP. 
    
    \item The runtime verification of the statistical parameters-based PSL assertion requires the hardware modules that compute the above-mentioned statistical parameters, i.e., the Hurst exponent ($H$), probability of hop distribution ($P$) and standard deviation of input injection distribution ($\sigma$). Therefore, during the design time, the designer designs these hardware modules and integrated with critical communication channels.  
    
\end{enumerate}

\textbf{During the run-time}, the values of the statistical parameters computed from the corresponding hardware modules are used to verify the associated PSL assertions. Note, for secure communication, and all the PSL assertions should be verified. If one of the assertions fails the verification then the communication channel is considered as intruded. After the verification failures, any of the suitable state-of-the-art recovery mechanisms can be applied, i.e., re-routing the communication, backup communication paths, backup components that are associated with critical computations or communication~\cite{cui2014high}.

\section{Hardware implementation}\label{sec:HW}
To extract the statistical parameters during the runtime for verification of the PSL assertions, we design the hardware modules for computing the Hurst exponent, hop distribution, and standard deviation. 
\begin{enumerate}

\item \textit{Hurst Exponent:} To compute the estimation, we used one of the most commonly used methods, i.e., the R/S method. The reason behind choosing this method is that it requires less number of observations to estimate the Hurst exponent.

    \begin{algorithm}[!t]{
			\footnotesize
			\caption{Hurst Exponent Estimation}
			\label{algo:hurst1}
			\begin{algorithmic}[1]
				\Statex \textbf{Input:}
				\State $n$: Number of the observations = 512
				\State $X[0:n-1]$ : number of communication packets per observation 
				\State Generate three computation blocks using $X[0:n-1]$
				
				\Statex \textbf{Computation Block 1:}
				\State Compute the mean and the standard deviation of $X[0:n-1]$
				\State Compute the mean centered series: $h[0:n-1]= X[0:n-1]-M$
				\State Compute cumulative deviation by summing up the mean centred values:
				\Statex $Y[0:n-1]=\sum_{i=0}^{0}h[i], \sum_{i=0}^{1}h[i], ..., \sum_{i=0}^{n-1}h[i]$
				\State Compute the Range (R) of $Y[0:n-1]$
				\State Compute $R/S$
				\State Compute $E(R/S)_0$
				
				\Statex \textbf{Computation Block 2:}
				\State Compute the mean (M) and standard deviation (S) of $X[0:(n/2)-1]$ and $X[n/2:n-1]$
				\State Compute the mean centered series: 
				\Statex $h_1[0:(n/2)-1]= X[0:(n/2)-1]-M$ and $h_2[n/2:n-1]= X[n/2:n-1]-M$
				\State Compute cumulative deviation by summing up the mean centred values:
				\Statex $Y[0:(n/2)-1]=\sum_{i=0}^{0}h_1[i], \sum_{i=0}^{1}h_1[i], ..., \sum_{i=0}^{(n/2)-1}h_1[i]$
				\Statex $Y[n/2:n-1]=\sum_{i=n/2}^{(n/2)}h_2[i], \sum_{i=n/2}^{(n/2)+1}h_2[i], ..., \sum_{i=n/2}^{n-1}h_2[i]$
				\State Compute the Range (R) of $Y[0:(n/2)-1]$ and $Y[n/2:n-1]$
				\State Compute $(R/S)_1$ and $(R/S)_2$
				\State Compute $E(R/S)_1 = \frac{(R/S)_1 + (R/S)_2}{2} $
				
				\Statex \textbf{Computation Block 3:}
				\State Compute the mean (M) and standard deviation (S) of $X[0:(n/4)-1]$, $X[n/4:(n/2)-1]$, $X[n/2:(3n/4)-1]$ and $X[3n/4:n-1]$
				\State Compute the mean centered series: 
				\Statex $h_1[0:(n/4)-1]= X[0:(n/4)-1]-M$, $h_2[n/4:(n/2)-1]= X[n/4:(n/2)-1]-M$, $h_3[n/2:(3n/4)-1]= X[n/2:(3n/4)-1]-M$ and $h_4[3n/4:n-1]= X[3n/4:n-1]-M$ 
				\State Compute cumulative deviation by summing up the mean centred values:
				\Statex $Y[0:(n/4)-1]=\sum_{i=0}^{0}h_1[i], \sum_{i=0}^{1}h_1[i], ..., \sum_{i=0}^{(n/4)-1}h_1[i]$
				
				\Statex $Y[n/4:(n/2)-1]=\sum_{i=n/4}^{n/4}h_2[i], \sum_{i=n/4}^{(n/4)+1}h_2[i], ..., \sum_{i=n/4}^{(n/2)-1}h_2[i]$
				
				\Statex $Y[n/2:(3n/4)-1]=\sum_{i=n/2}^{n/2}h_3[i], \sum_{i=n/2}^{(n/2)+1}h_3[i], ..., \sum_{i=n/2}^{(3n/4)-1}h_3[i]$
				
				\Statex $Y[3n/4:n-1]=\sum_{i=3n/4}^{3n/4}h_4[i], \sum_{i=3n/4}^{(3n/4)+1}h_4[i], ..., \sum_{i=3n/4}^{n-1}h_4[i]$
				\State Compute the Range (R) of $Y[0:(n/4)-1]$, $Y[n/4:(n/2)-1]$, $Y[n/2:(3n/4)-1]$ and $Y[3n/4:n-1]$
				\State Compute $(R/S)_1$, $(R/S)_2$, $(R/S)_3$ and $(R/S)_4$
				\State Compute $E(R/S)_2 = \frac{(R/S)_1 + (R/S)_2 + (R/S)_3 + (R/S)_4}{4} $
				
				\Statex \textbf{Hurst Exponent:}
				\State Compute $E(R/S) = \frac{E(R/S)_0 + E(R/S)_1 + E(R/S)_2}{3} $
				\State Compute Hurst exponent $H = 0.37 \times log{E(R/S)}$ 
		\end{algorithmic}}
	\end{algorithm}

\begin{figure}[!t]
	\centering
	\includegraphics[width=1\textwidth]{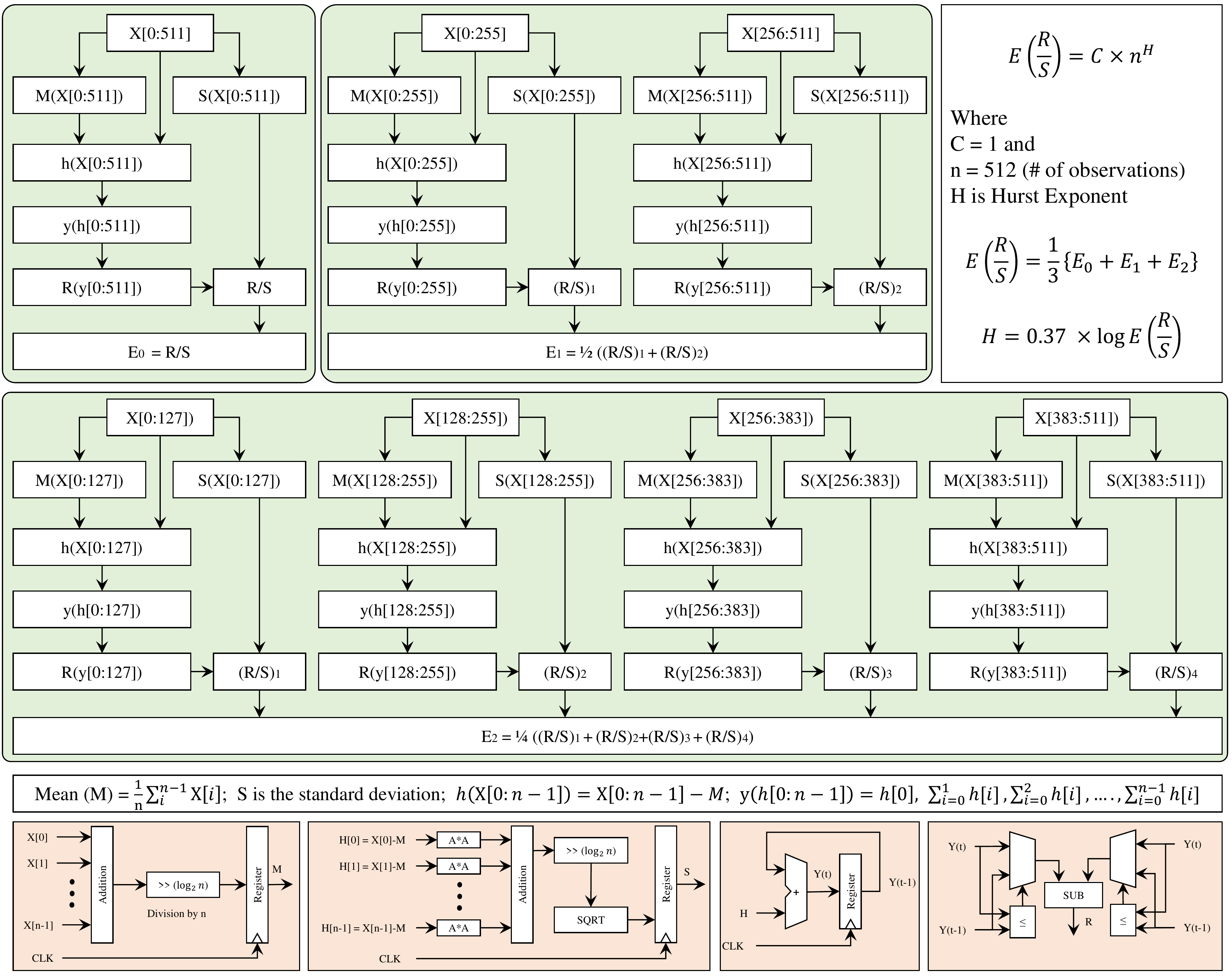}
	\caption{Data flow of the hardware module to compute the Hurst exponent. In this paper, we set the number of observations to 512. The reason behind choosing this value is that for fast convergence of Hurst exponent, the number of observations should be greater than 300. Note, To estimate the Hurst exponent, we used one of the most commonly used methods, i.e., the R/S method. The reason behind choosing this method is that it requires fewer number of observations to estimate the Hurst exponent. We implemented state-of-the-art modified non-restoring algorithm~\cite{sutikno2012simplified} for computing the square root function in standard deviation.}
	\label{fig:HW}
\end{figure}

\begin{itemize}
    \item After establishing a communication channel, it observes and stores the number of the packets per clock cycle, as denoted by $X$ in Algorithm~\ref{algo:hurst1}.
            
    \item To estimate the Hurst exponent, it is important to take the average value of R/S values using the different data distribution obtained from the same data. Therefore, we propose to use three computational blocks, as shown in Algorithm~\ref{algo:hurst1} and Figure~\ref{fig:HW}. The data flow and hardware modules in each computational block are the same but the sizes of hardware modules are different. The computational block 1 uses the complete data for estimating the R/S value. The computational block 2 divides the data into two equal parts. Then this block estimates the respective R/S values using each half data and takes the average to estimate the R/S value. Similarly, the computational block 3 repeats the same procedure, but it divides the data into four equal parts. 
            
    \item In each computational block, the first step is to compute the mean value (M) and standard deviation (S) of the input data series $X$. The hardware modules to compute the mean value consists of ``n’’ 64-bit adders, one 9-bit shifter and one output register, as shown in Figure~\ref{fig:HW}. The hardware module for computing the standard deviation consists of ``n’’ 64-bit subtractors, ``n’’ 64-bit multipliers, one 9-bit shifter, one module to compute the square root and one output register. Note, we implemented state-of-the-art modified non-restoring algorithm~\cite{sutikno2012simplified} for computing the square root function in standard deviation.
            
    \item After computing the mean and standard deviation, each computational block computes the mean-centered data series ($H$) by subtracting the mean value from each input data series ($X$), as shown in Algorithm~\ref{algo:hurst1}. Then, it computes the cumulative deviation ($Y$) by summing up the mean-centered data series ($H$) and computes the magnitude range ($R$) of the cumulative deviation ($Y$). Finally, it computes the R/S using a 64-bit divider. Note, there is no extra hardware for the mean-centered series and each computational block uses the values from the standard deviation module. The hardware module of computing $Y$ consists of one 64-bit accumulator, as shown in Figure~\ref{fig:HW}, and the hardware module computing $R$ consists of two comparators, two multiplexers and one 64-bit subtractor. 
            
    \item Finally, it computes the R/S by taking the average of R/S values computed from each computational block. Finally, the average R/S value is used to compute the Hurst exponent using $H = 0.37\times log_{10} (average value of R/S)$, as shown in Algorithm~\ref{algo:hurst1} and Figure~\ref{fig:HW}.   

\end{itemize}

\item \textit{Standard Deviation}
We have not used a separate block for computing the standard deviation. We use the intermediate output from the standard deviation block of Hurst exponent.
        
\item \textit{Hop Probability }
In order to compute the hop probability, SIMCom computes the number of active channels by counting the the acknowledgment signals during the channel establishment. Note, this parameter is effective in the case of denial-of-service, jamming or communication blocking.

\end{enumerate}
\section{Case Studies}\label{case-study}

To illustrate the scalability of the SIMCom, we implemented the three SoCs, as shown in Fig\ref{fig:exp_setup}.
\begin{enumerate}
    \item \textbf{SoC1} consists of four single-core MC8051 with UART modules.
    \item \textbf{SoC2} consists of four single-core MC8051 linked with each other and AES, ethernet, memctrl, BasicRSA, RS23s modules.
    \item \textbf{SoC3} consists of four single-core LEON3 connected with each other and AES, ethernet, memctrl, BasicRSA, RS23s modules.
\end{enumerate}
All the SoC1 are tested on the Trust-Hub HT benchmarks~\cite{trust-HUB}. Note for SoC1, and the input data distribution is Gaussian and exponential. The workloads used for LEON3 in SoC3 are 64-bit encrypted multiplication, subtraction, addition, and division with randomly distributed input data. The input data is encrypted using AES. Results outputted using VGA display, ethernet, and RS232. For MC8051 in SoC2, we use the same set of applications with 32-bit. For area and power overhead analysis, we synthesized the SoCs using Cadence Genus (Encounter) tool with the TSMC 65nm library.  

\subsection{SoC1: Four MC8051 IPs are connected with UART IPs}

\begin{figure}[!t]
	\centering
	\includegraphics[width=1\linewidth]{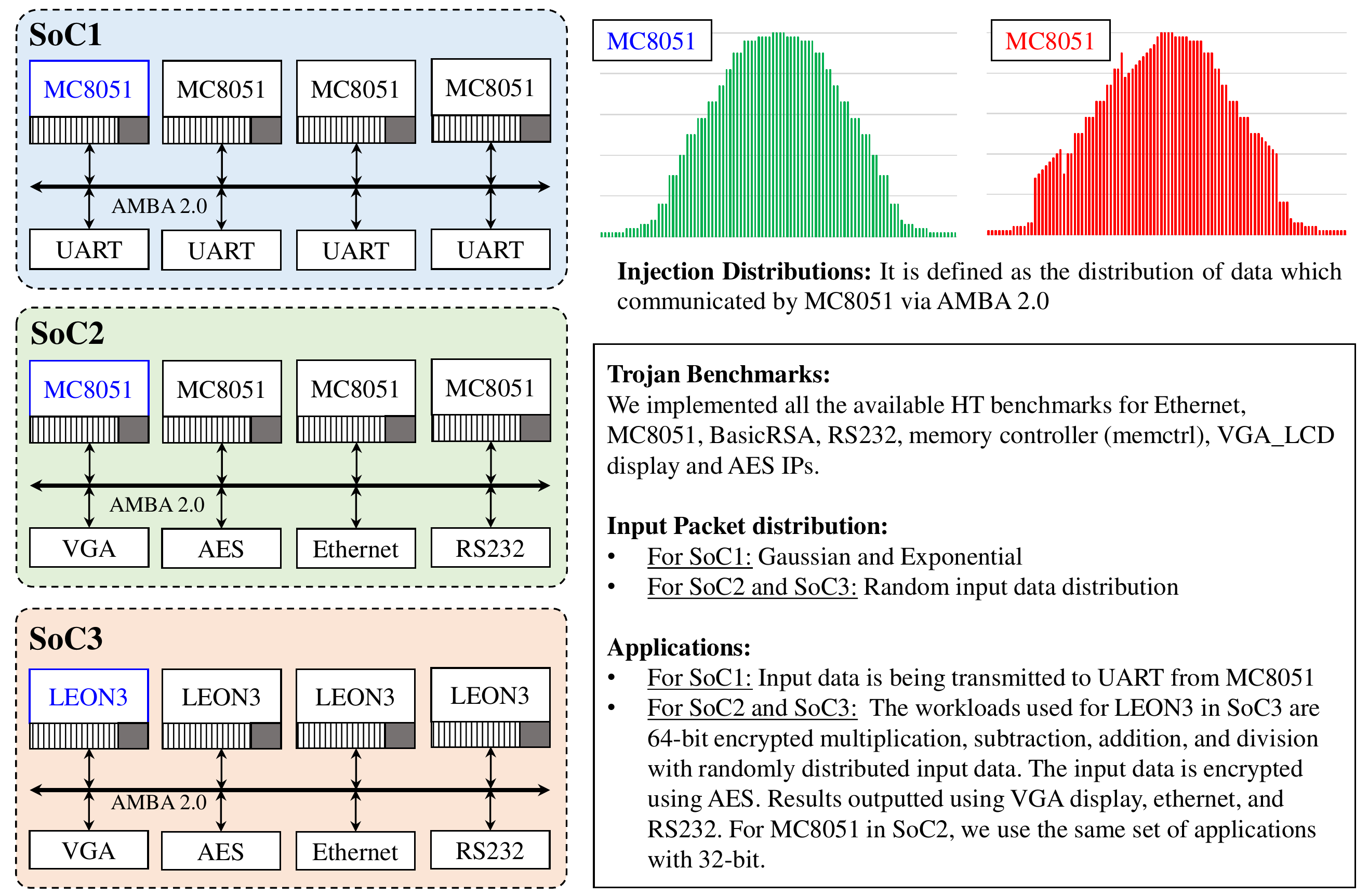}
	\caption{Experimental Setup and a brief overview of all the implemented SoCs.}
	\label{fig:exp_setup}
\end{figure}

In this section, we illustrate the effectiveness of SIMCom by applying it on the SoC with AMBA bus connecting an MC8051 microcontroller and a UART module, as shown in Figure \ref{fig:exp_setup}.  
This experimental setup consists of four MC8051 Master modules, which can initiate the communication with the eight slave modules along with two UART modules. The main motivation of choosing the MC8051 microcontroller as our case study is the availability of its open-source Trojan benchmarks at the trusthub.org \cite{trust-HUB}. We first obtained the communication behavior of the \textit{un-intruded MC8051} while communicating with the UART modules. We used this behavior to obtain the corresponding PSL assertions and embed them into the SoC. The key assumption of the experimental setup is that \textit{Though all instances of the MC8051 can have the Trojan but Trojan in only one IP module is triggered at a particular time considering a unique sequence happening at run time with a very low probability of triggering (almost zero)}. 

\subsection{Pre-Market Test Stage Traffic Model of MC8051} \label{TM-8051_model}
For obtaining this behavior, we implemented the MC8051 microcontroller in VHDL and synthesized it for the Spartan 6 (xc6slx45)\footnote{Since the overall SoC fits into this small FPGA, there was no need to use a bigger FPGA chip. We target a low-cost SoC, for which Spartan 6 suffices the need.} FPGA. The communication behavior is extracted by measuring the number of packets for the specific duration and computing the required traffic model parameters, $H$, $P$ and $\sigma$. This phase of our methodology is divided into the following steps:
\begin{enumerate}
	\item The first step is to extract the communication model. We developed the model by injecting the packets using the Gaussian distribution, as shown in Figure \ref{fig:exp_setup}. This probability distribution not only reflects the actual traffic very well, but it also incorporates the environmental and process variations \cite{kreutz2001communication}. In our experimental setup, all the master modules, i.e., MC8051s, inject packets into the SoC. The value of $\sigma$ of MC8051 for the first 10,000 clock cycles is kept as 18.844 and maintained over the 100,000 clock cycles, as shown in Figure~\ref{fig:results_time}.
	\item Afterwards, we need to identify the number of modules that a particular master can access through the bus. In this case study, we assume that every MC8051 can communicate with all the slave modules ($n = 8$). However, for some specific instruction sets, MC8051 does not communicate with all the slave modules. For example, the instructions that require the UART communication module only. In our setup, there are only \textit{``two''} slaves that are using the UART communication modules. Therefore, the acceptance probability ($p$) for this case study must be greater than 0.125 as shown in Figure \ref{fig:results_time}. Its average value over the 100,000 clock cycles is indicated in column ``$P$'' under ``T000 (Without Trojan)'' of Table \ref{tab:tab2}. \par
	\begin{algorithm}[!t]{
			\footnotesize
			\caption{Hurst Exponent Estimation in SoC consists of MC8051 and UART}
			\label{algo:hurst}
			\begin{algorithmic}[1]
				\Statex \textbf{Input:}
				\Statex \tab $b_{size}$: Minimum Number of Clock Cycles (Bin Size)
				\Statex \tab $max_{clk}$: Maximum Number of Clock Cycles
				\Statex \tab $np$: Number of packet per bin; $np[0:99]$: Number of packet per bin
				\Statex \tab $H$: Hurst Exponent; $H[0:99]$: Hurst parameter array
				\Statex \textbf{Initialize:}
				\Statex \tab $bin_{size}$: 99;  $max_{size}$: 99;  $n_{packets}$: 0; $i$: 0; $j$: 0;$k$: 0;
				\While{$k \leq 10$} 
				\While{$i \leq max_{clk}$} 
				\While{$j \leq b_{size}$}
				\State $j = j + 1\ @\ posedge\ clock$; $np = np + 1 \ @\ each\ packet$
				\EndWhile
				\State $np[i] = np $; $i = i + 1$
				\If {$np[i] \geq np[i-1]$}
				\State	$np_{max} = np[i]$ //Maximum number of packets per Bin
				\Else 
				\State	$np_{max} = np[i-1]$	
				\EndIf
				\If {$np[i] \leq np[i-1]$}
				\State	$np_{min} = np[i]$ //Minimum number of packets per Bin
				\Else 
				\State	$np_{max} = np[i-1]$	
				\EndIf
				\State $np_{total} = np[i] + np[i+1]$; $j = 0$ 
				\EndWhile
				\State $np_{average} = np_{max}/i$				
				\State $H[k] = \log_{10}(np_{max} - np_{min})/(np_{average}\times\log_{10} (max_{clk}. b_{size}))$
				\State $i = 0$; $H_{total} = H[k] + H[k-1]$
				\EndWhile
				\State $H = H_{total}/k $	
		\end{algorithmic}}
	\end{algorithm}	

	\item The last step in obtaining the communication model is to estimate the Hurst exponent for MC8051. For this, we have developed Algorithm \ref{algo:hurst}, which counts the number of packets for a specific duration and estimates the Hurst exponent ($H$). In Algorithm \ref{algo:hurst}, a counter ($np$) is used to count the number of packets for a specific bin size. For the considered study, its value is 100 clock cycles. The counting is done 100 times to find the maximum ($np_{max}$) and minimum ($np_{min}$) number of packets in 100 clock cycles. However, for a larger number of clock cycles, the memory requirement for storing these parameters is very high. Therefore, to reduce the memory requirement, we divided the whole duration into 100 equal parts. For example, in Algorithm \ref{algo:hurst}, the total time duration is 10,000 clock cycles, which is divided into 100 smaller time duration part. For each duration, the number of packets is calculated, which is stored in a variable array $np[0:99]$ to compute its overall range ($np_{max} - np_{min}$) and average ($\dfrac{np_0 + np_1 + .. + np_{99}}{100}$). Finally, the Hurst exponent is computed by averaging out the obtained values. We repeated these steps 10 times, i.e., data of the 10,000 clock cycles data is stored in one array and then the whole process is repeated 9 times to get an overall $H$ exponent for 100,000 clock cycles. This value is tabulated as 0.692 in the column ``$H$'' under ``T000'' (without Trojan) of Table \ref{tab:tab2}.   \par   
\end{enumerate}
	
\subsection{PSL Assertions}\label{PSL}
We analyzed the extracted pre-market test stage communication model  of the AMBA bus, and translated it into the following PSL assertions based on the traffic model parameters, $H$, $P$ and $\sigma$:
\begin{enumerate}
	\item \textbf{P Assertion:} It states that, over a given number of clock cycles, the acceptance probability ($P$) for a particular module must be equal to the value of acceptance probability ($P$) to the obtained communication behavior: 
	\begin{align*}	
	\texttt{\textbf{always} (P[1:100K] == P\_design)} \tag{I}\label{prop:1} 
	\end{align*}
	Moreover, the acceptance probability ($P$) for a particular architecture must be within the prescribed limits. The maximum and the minimum values of $P$ are obtained as ($P_{max} = \dfrac{1}{n-1}$) and ($P_{min} =\dfrac{1}{r-1}$), respectively, where $n$ and $r$ represent the number of modules that each module can communicate with, and the number of modules that a particular instruction set/application need to communicate with, respectively. This is elaborated in Assertion \ref{prop:2}:  
	\begin{align*}	
	\texttt{\textbf{always} (P[1:100K] <= Pmax\ \&\& P[1:100K] >= Pmin)} \tag{II}\label{prop:2} 
	\end{align*}
	\item \textbf{$\sigma$ Assertion:} This assertion states that over a given number of clock cycles, the value of $\sigma$ for a particular module must be equal to the value of $\sigma$ from the pre-market test stage communication behavior:
	\begin{align*}	
	\texttt{ \textbf{always} (sigma[1:100K]== sigma\_design)} \tag{III}\label{prop:3} 
	\end{align*} 
	\item \textbf{H Assertion:} This assertion ensures that over a give number of clock cycles, the Hurst exponent ($H$) for a particular module must be equal to value of $H$ obtained at the pre-market test stage:		
	\begin{align*}	
	\displaystyle \texttt{ \textbf{always} (H[1:100K]== H\_design)} \tag{IV}\label{prop:4} 
	\end{align*}
\end{enumerate}
In state-of-the-art HT detection techniques for SoC and microprocessors, the process variation margin is considered as approximately 10\%, however, this margin depends upon the technology parameters~\cite{wang2008detecting,brasser2018special}. Therefore, to incorporate the environmental changes and process variation, we proposed to introduce a 10\% variation cap on the $H$, $P$, $\sigma$ based on the analysis presented in~\cite{gil2012studying,korak2014effects}.
\begin{table}[!t]
	\centering
	\caption{Effects of the trust-hub Trojan benchmarks on different modules of MC8051 and corresponding statistical modeling. $\times$ and $\checkmark$ shows that Trojan either affect or does not affect a particular functionality.}
	\label{tab:tab2}
	\resizebox{0.7\textwidth}{!}{
		\begin{tabular}{|c|c|c|c|c|}
			\hline
			\textbf{Trojans} & \textbf{\begin{tabular}[c]{@{}c@{}}UART\end{tabular}} & \textbf{\begin{tabular}[c]{@{}c@{}}Jump \\ Disabling\end{tabular}} & \textbf{\begin{tabular}[c]{@{}c@{}}ALU \\ Operations\end{tabular}} & \textbf{($H$,$\sigma$,$P$ )} \\ \hline
			\textbf{T000} & -- & -- & -- & (0.694,18.844, 0.252)\\ \hline
			\textbf{T200} & -- & -- & -- & --\\ \hline
			\textbf{T300} & $\checkmark$ & $\times$ & $\checkmark$ & (0,0,0)\\ \hline
			\textbf{T400} & $\times$ & $\times$ & $\checkmark$ & (0.704,14.617, 0.252)\\ \hline
			\textbf{T500} & $\checkmark$ & $\times$ & $\checkmark$ & (0.701,15.019, 0.252)\\ \hline
			\textbf{T600} & $\checkmark$ & $\checkmark$ & $\checkmark$ & (0.644,14.472, 0.252) \\ \hline
			\textbf{T700} & $\checkmark$ & $\times$ & $\times$ & (0.672,14.728, 0.252)\\ \hline
			\textbf{T800} & $\checkmark$ & $\times$ & $\times$ & (0.639,14.476, 0.252) \\ \hline
	\end{tabular}}
\end{table}
\subsection{Run-time Detection and Validation}\label{PSL_run}
The last step of the proposed methodology is to embed the translated assertions along with the $H$, $P$ and $\sigma$ estimation blocks into the SoC. To illustrate the proposed methodology, we validated it with the MC8051 benchmarks intrusions, which are available online at trust-hub.org \cite{trust-HUB}. These intrusions affect different types of communication behavior. Some of them affect the UART communication, other affect the jump based instruction and some intrusions affect the ALU operations, as given in Table \ref{tab:tab2}. Each row in Table \ref{tab:tab2} represents the benchmark Trojans, and the columns describe their effects on different operations. We validated the proposed methodology for the intrusions T300 to T800, because these are all the intrusions that can directly or indirectly affect the communication of MC8051 among the set of the benchmark Trojans. To evaluate the consistency checking of SIMCom, we analyzed the runitme and average statistical behavior for the experimental setup. 

\subsubsection{Run-time Impact Analysis} 
\begin{figure}[!t]
		\centering
		\includegraphics[width=1\linewidth]{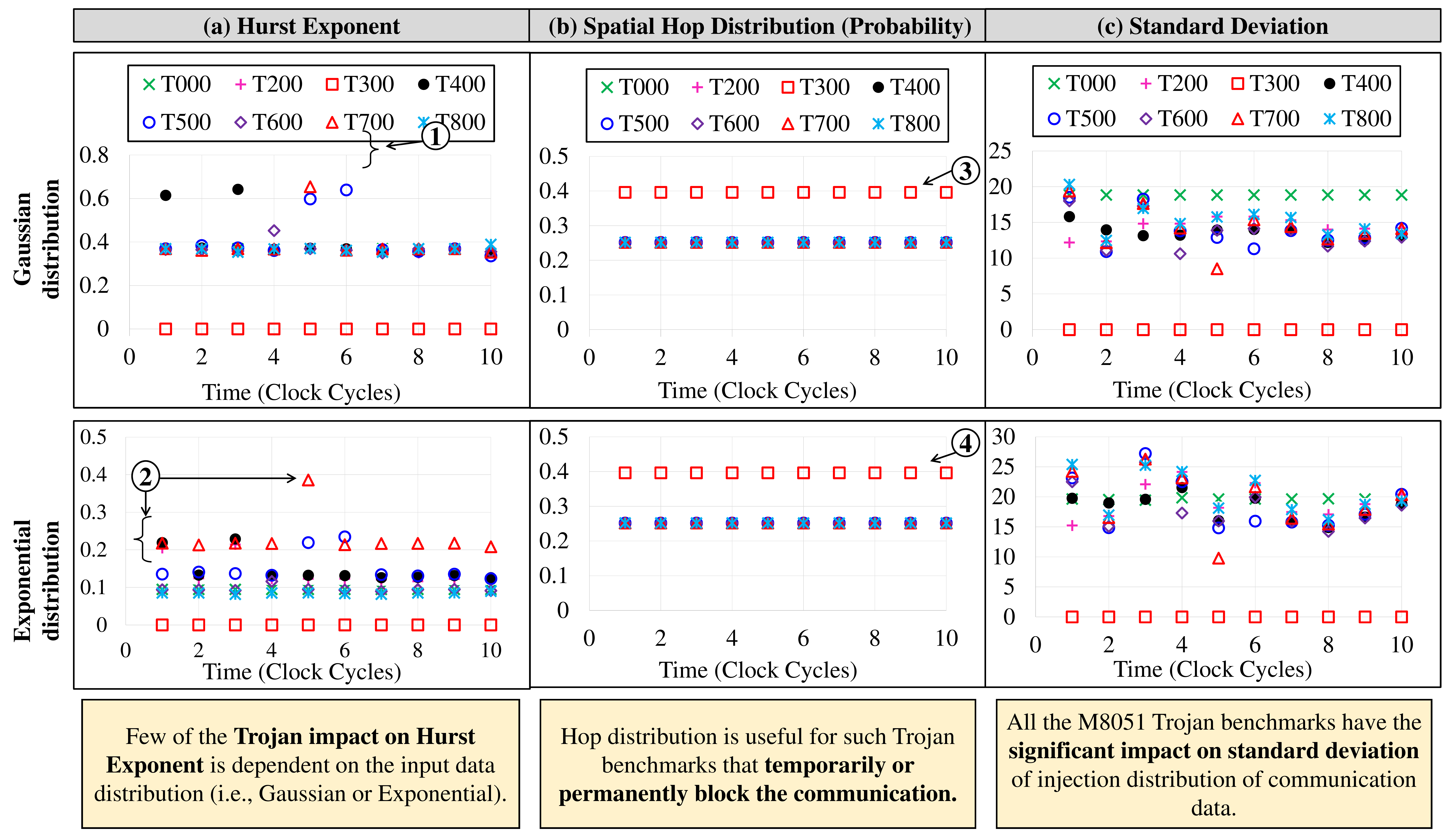} 
		\caption{Runtime Impact Analysis of implemented HTs (i.e., MC8051-T200, T300, T400, T500, T600, T700, T800) on the statistical model ($H$, $P$, $\sigma$) of the implemented case study for 100,000 clock cycles and the values are averaged out for 10,000 clock cycle duration. Labels 1, 2, 3 and 4 of this figure is described in Section \ref{TM-8051_model}.}
		\label{fig:results_time}
		\vspace{0.1in}
	\end{figure}

To elaborate on the practicality of SIMCom, we have computed the statistical communication behavior of the case study for 100,000 clock cycles, as shown in Figure \ref{fig:results_time}. The figure depicts the communication behavior with respect to $H$, $\sigma$ and $P$. This run-time analysis exhibits the following key observations:\par
\begin{enumerate}
	\item \textbf{Hurst Exponent:} The Trojan benchmarks exhibit a significant impact on the Hurst Exponent depending on the input data distribution (i.e., Gaussian or Exponential). For instance, in Figure \ref{fig:results_time}, if the input data distribution is\textit{ Gaussian}, then a few of the implemented Trojan benchmarks (i.e., MS8051-T400, T500, and T700) have a significant impact on the Hurst exponent (see: label 1). However, if the input data distribution is \textit{exponential}, then almost all the HTs exhibit a significant impact on the Hurst exponent (see: label 2).
	\item \textbf{Probability of Hop Distribution:} All the implemented Trojans exhibit no impact on the probability of hop distribution, except the MC8051-T300 because when it is triggered, it halts a communication channel, as shown in Figure \ref{fig:results_time} (see: labels 3 and 4). 
	\item \textbf{Standard Deviation of Injection Distribution:} All the implemented Trojans deviate from the original values because all of them affect the input data injection, as shown in Figure \ref{fig:results_time}. For instance, MS8051-T500 and T700 replace valid data with intruded data. However, MC8051-T400 disables \textit{interrupt handling}, T600 \textit{interrupts the jump} and T800 \textit{manipulates the stack} pointer, which indirectly disrupts the input data or respective control modules.    	
\end{enumerate}
The values of $H$ and $\sigma$ for MC8051-T300 are ``0'' because upon triggering T300 blocks the UART communication altogether, and \textit{hence no data traffic flows at all}, resulting in the corresponding '0' values for \textit{H} and sigma. \par   

\subsubsection{Average Impact Analysis}
\begin{figure}[!t]
    	\centering
    	\includegraphics[width=1\linewidth]{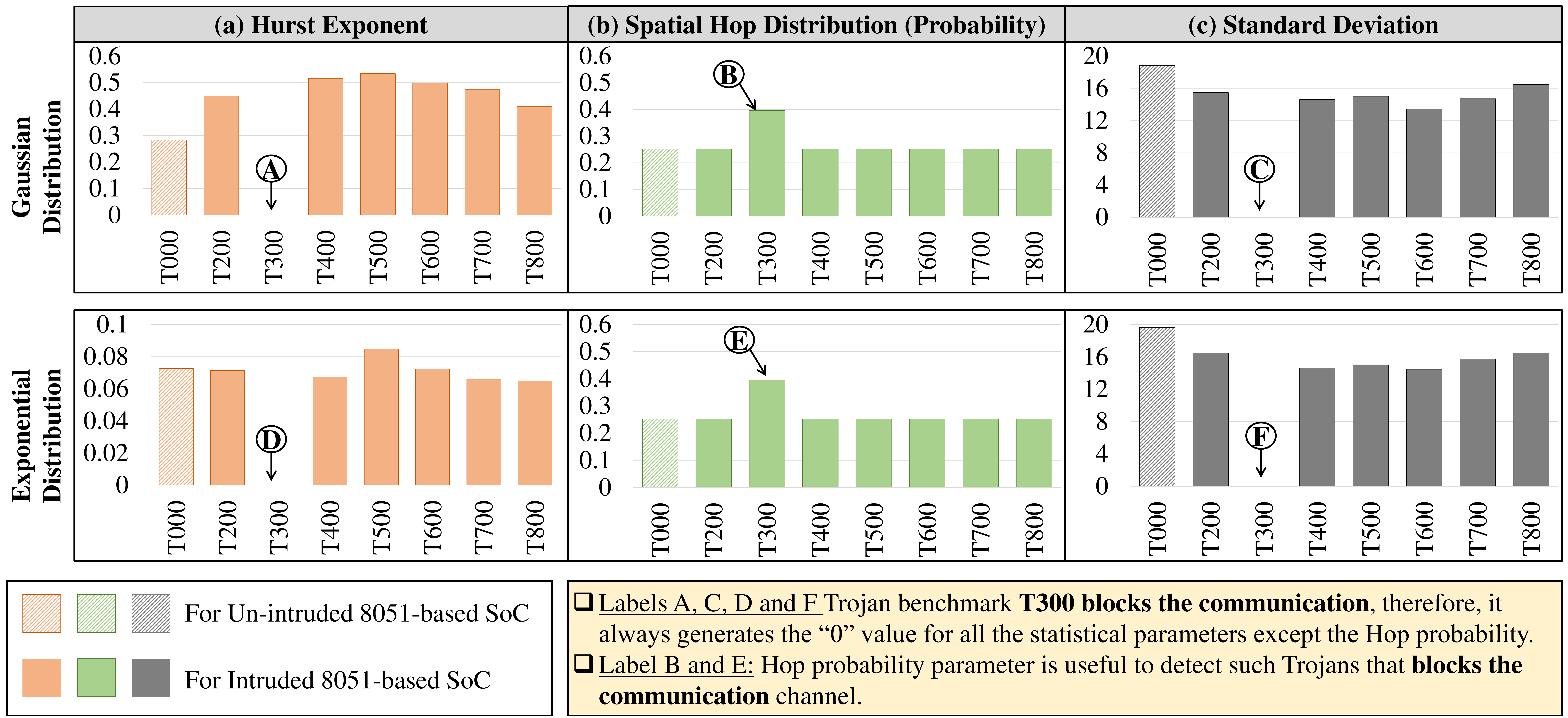} 
    	\caption{Average Impact Analysis of implemented Trojans (i.e., MC8051-T200, T300, T400, T500, T600, T700, T800) statistical model of 8051 based UART communication network for 100000 clock cycles. Note: T000 represents the pre-market test stage communication behavior of the MC8051 based experimental case study  with no Trojans inserted.}
    	\label{fig:results_avg}
    	\vspace{0.1in}
    \end{figure}

In order to elaborate the consistency of SIMCom, we have analyzed the average communication behavior of the MC8051-based UART communication network, as shown in Figure \ref{fig:results_avg}, which depicts the communication behavior with respect to $H$, $\sigma$ and $P$. This analysis exhibits the following key observations:\par
\begin{enumerate}
	\item \textbf{Hurst Exponent:} The average behavior of the proposed statistical model shows that, on the average, all the implemented Trojan benchmarks have a significant impact on the $H$ irrespective of the input data distribution, as shown in Figure \ref{fig:results_avg}. However, the value of $H$ for MC8051-T300 is ``0''. Hence, this can be considered as the weakest Trojan for the communication-aware HT detection (see: label A and D). Though it shows that the run-time impact of implemented Trojans is more in case of the exponential distribution, in average analysis, this impact is less as compared to the impact on the Gaussian distribution. The reason behind this behavior is that the data rate decreases exponentially over any period of time in the exponential distribution.
	\item \textbf{Probability of Hop Distribution:} Even in the average-case analysis, all the implemented Trojans exhibit no impact on the probability of hop distribution, except for the MC8051-T300 as shown in Figure \ref{fig:results_avg} (see: labels D and E). The reason behind this behavior is the blockage of the communication of a channel when this Trojan benchmark gets an activation trigger.  Hence this parameter can be used for detecting such (T300) kinds of Trojans, which are otherwise difficult to be detected.   
	\item \textbf{Standard Deviation of Injection Distribution:} All the implemented Trojans deviate from the original values because all of them affect the input data injection, except the MC8051-T300, as shown in Figure~\ref{fig:results_avg} (see: labels C and F). For instance, MS8051-T500 and T700 replace valid data with the intruded data. However, the MC8051-T400 disables \textit{interrupt handling}, the T600 \textit{interrupts the jump} and the T800 \textit{manipulates the stack} pointer, which indirectly disrupts the input data or respective control modules.    	
\end{enumerate}

\subsection{Experimental Analysis for SoC2 and SoC3}
\begin{figure}[!t]
        \centering
        \includegraphics[width=1\linewidth]{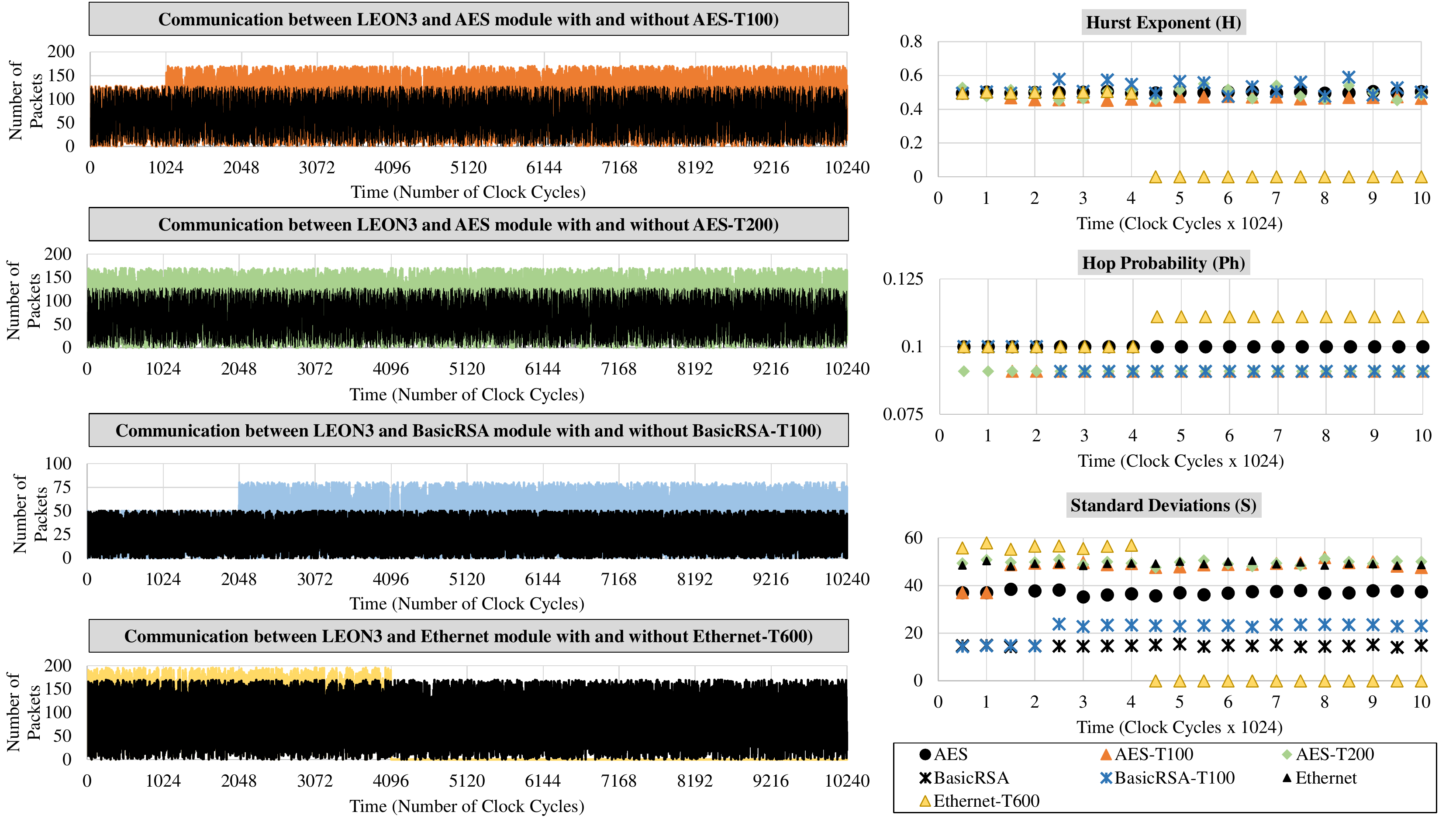} 
        \caption{Experimental results to show the effect of HT benchmarks, i.e., AES-T100, AES-T200, BasicRSA-T100 and EthernetMAC10GE-T600, on communication between \textbf{MC8051} and AES, ethernet and basic RAS module respectively. Note, in this analysis the number of the observations (n) is 512.}
    	\label{fig:result_MC8051}
    \end{figure}

\begin{figure}[!t]
    	\centering
    	\includegraphics[width=1\linewidth]{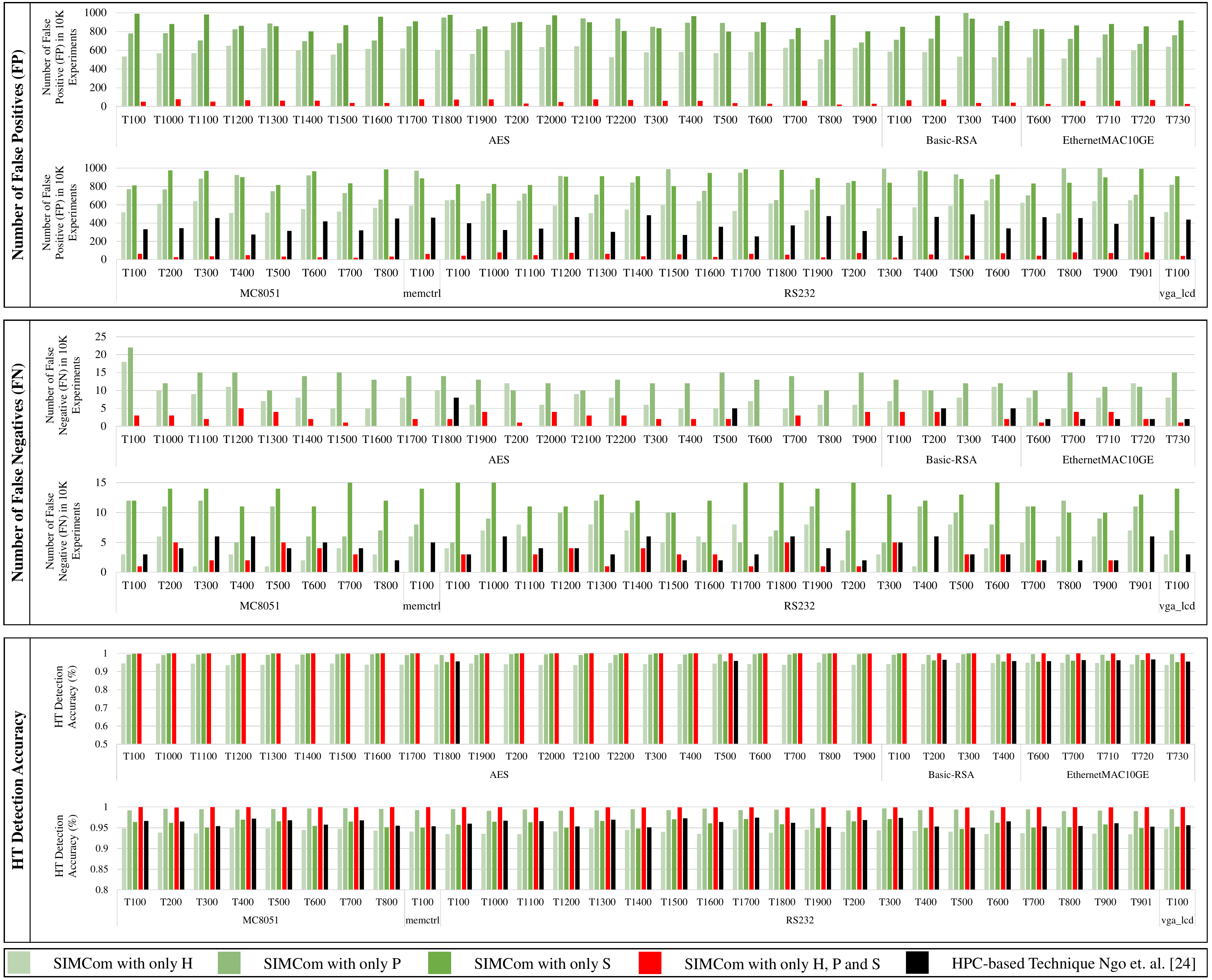} 
    	\caption{False positive, False-negative and HT detection accuracy of the SIMCom and the state-of-the-art run-time detection technique~\cite{ngo2015hardware}, in the presence of different HT benchmarks, on \textbf{MC8051-based SoC}. Note, these analyses are based on the 10000 observations, where the total number of HT activations is 1000. In these experiments, the overall accuracy is computed as $Accuracy = \frac{TP+TN}{TP+TN+FP+FN}$, where TP, TN, FP, and FN represent true-positives, true-negatives, false-positives, and false-negatives, respectively.} 
    	\label{fig:FP_FN_8051}
    \end{figure}

\begin{figure}[!t]
        \centering
        \includegraphics[width=1\linewidth]{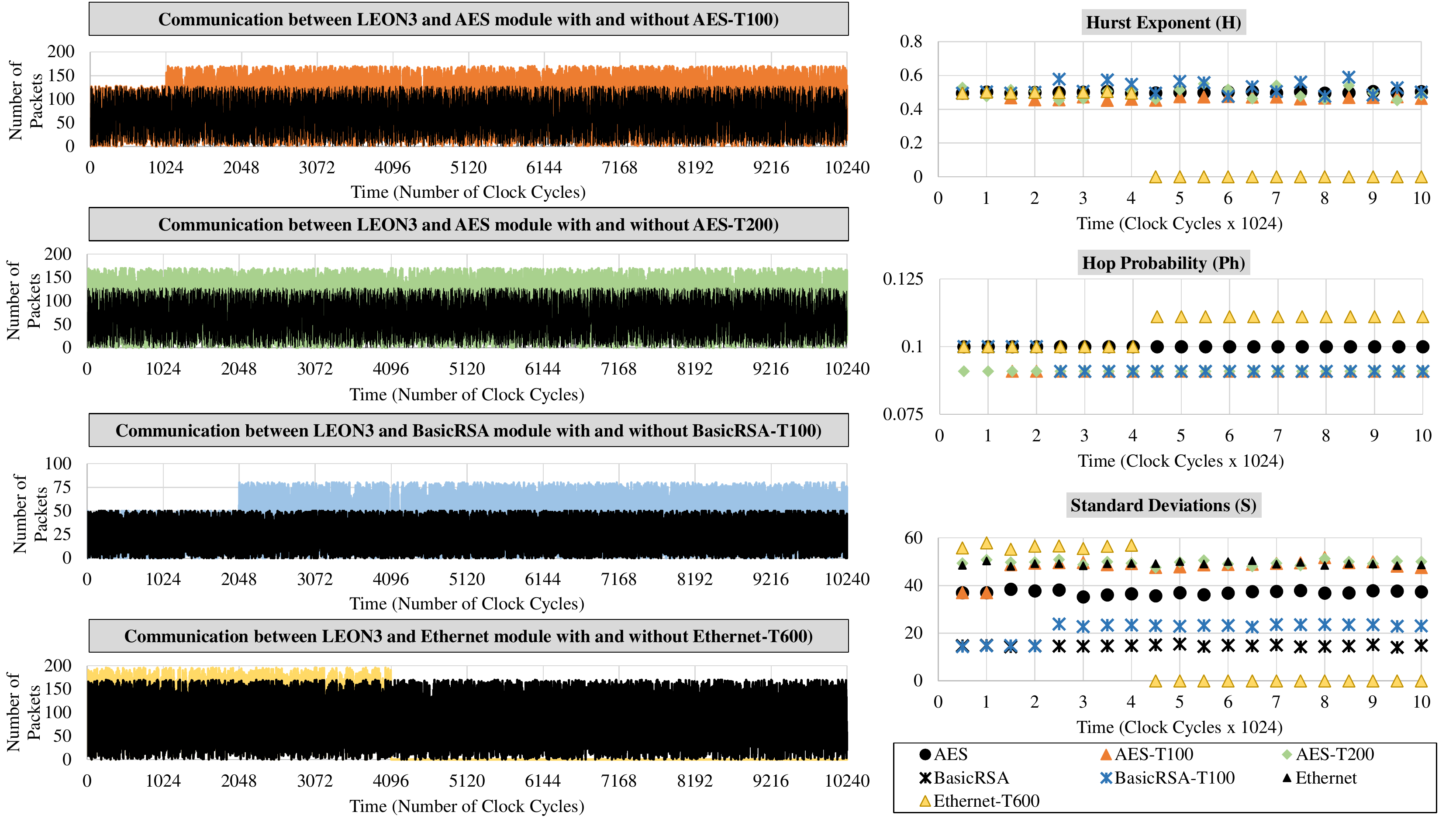} 
        \caption{Experimental results to show the effect of HT benchmarks, i.e., AES-T100, AES-T200, BasicRSA-T200 and EthernetMAC10GE-T600, on communication between \textbf{LEON3} and AES, ethernet and basic RAS module respectively. Note, in this analysis the number of the observations (n) is 512.}
    	\label{fig:result_LEON3}
    \end{figure}

\begin{figure}[!t]
    	\centering
    	\includegraphics[width=1\linewidth]{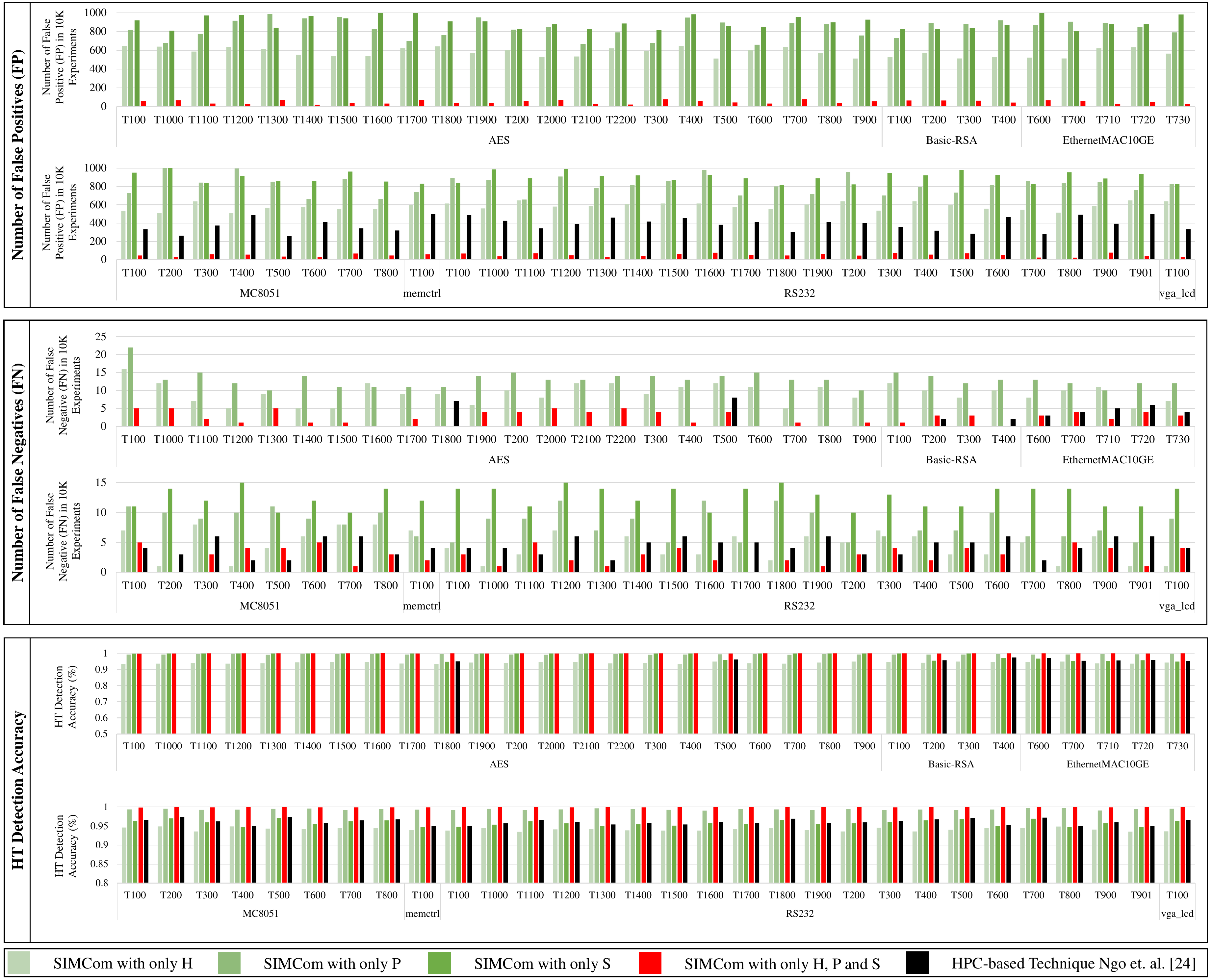} 
    	\caption{False positive, False-negative and HT detection accuracy of the SIMCom and the state-of-the-art run-time detection technique~\cite{ngo2015hardware}, in the presence of different HT benchmarks, on \textbf{LEON3-based SoC}. Note, these analyses are based on the 10000 clock cycles, where the total number of HT activations is 1000. In these experiments, the overall accuracy is computed as $Accuracy = \frac{TP+TN}{TP+TN+FP+FN}$, where TP, TN, FP, and FN represent true-positives, true-negatives, false-positives, and false-negatives, respectively.} 
    	\label{fig:FP_FN_LEON3}
    \end{figure}

We analyze timing behavior of the communication pattern, false positives, false negatives and HT detection accuracy in SoC2 and SoC3.
\begin{itemize}
    \item \textbf{Timing behavior of H, P, and S:}  Figures~\ref{fig:result_MC8051}~and~\ref{fig:result_LEON3} show the timing behavior of Hurst exponent, standard deviation, and hop probability for three communication channels, i.e., MC8051/LEON3 with AES, BasicRSA, and Ethernet, in the presence of four HT benchmarks, i.e., AES-T100, AES-T200, BasicRSA-T100, and EthernetMAC10GE-T600. This analysis shows the significant change in H and S. Note, only the communication blocking affects the hop probability. Therefore, by analyzing all these parameters, SIMCom detects HTs and its payload type. For example, if it affects the P, then the payload is either denial-of-service, communication blocking or jamming. 
    
    \item \textbf{HT Detection Accuracy:} Figures~\ref{fig:FP_FN_8051}~and~\ref{fig:FP_FN_LEON3} show the false positives, false negatives and HT detection accuracy of the different configurations of SIMCom, i.e., SIMCom with only H, SIMCom with only P, SIMCom with only S, and SIMCom with H, P and S. The analysis shows that HT detection accuracy of SIMCom with H, P and S is approximate, 99\%. Note, SIMCom with the only P detects only those Trojans that blocks the communication channels like EthernetMAC10GE-T600.    
\end{itemize}
  
\subsection{Overhead Analysis}
\begin{figure}[!t]
        \centering
        \includegraphics[width=1\linewidth]{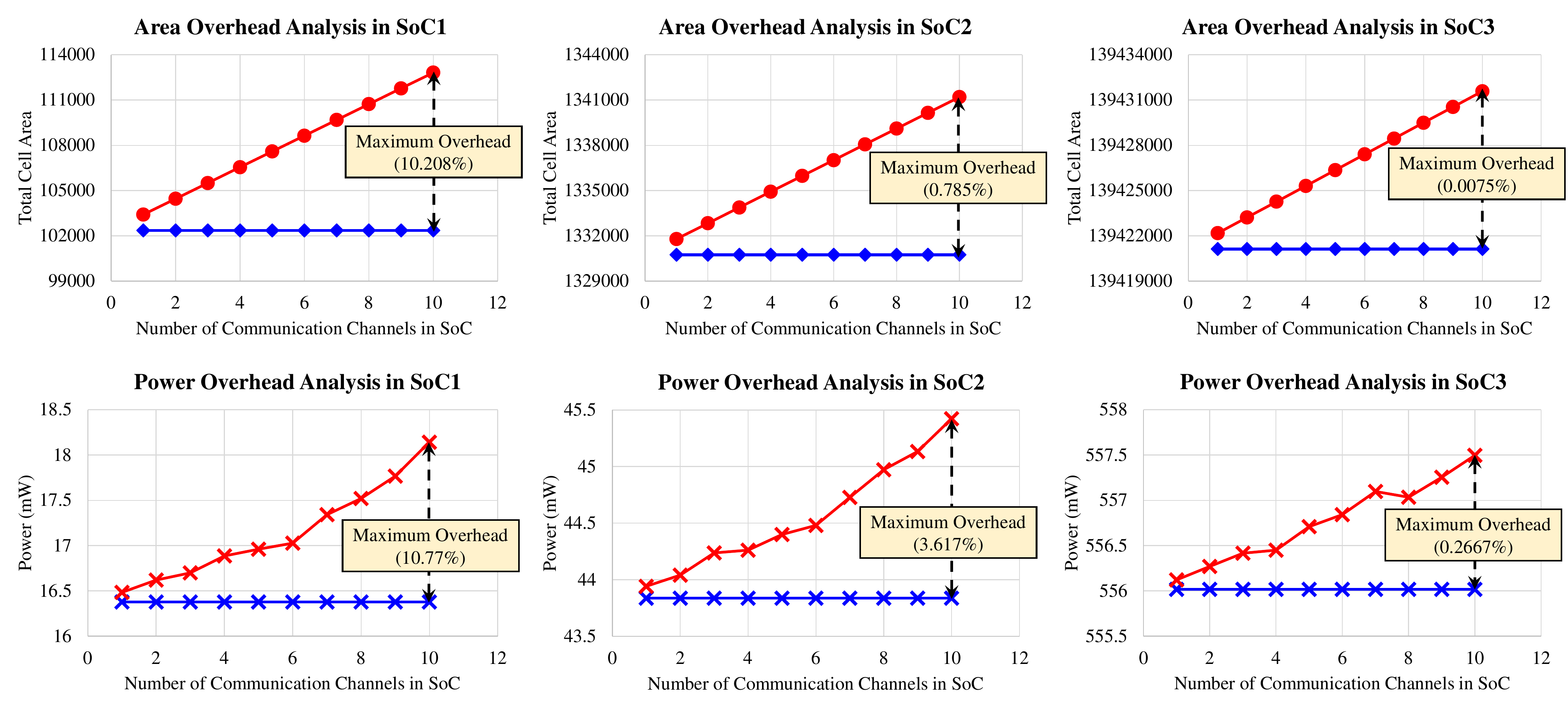} 
        \caption{Area and power overhead analysis of implemented SoCs.}
    	\label{fig:overhead}
    \end{figure}

For area and power overhead analysis, we synthesized the SoCs using Cadence Genus (Encounter) tool with the TSMC 65nm library. The overhead analysis presented in Figure~\ref{fig:overhead} shows the following observations: 

\begin{enumerate}
    \item The area and power overhead associated with SIMCom in the SoCs with the same number of communication channels relatively decreases with the increase in the complexity of the modules. For example, in SoC3, the area and power overhead is less than 1\%.
    
    \item If the number of communication channels increases, the overhead of the SIMCom also increases. Therefore, we propose to use the state-of-the-art methodology for distributing the runtime monitors in the SoC~\cite{khalid2018runtime}. This distribution technique proposes to three topologies, i.e., a global monitor, region-based distribution, and channel-based monitors, that can be selected based on the power and area budget. 
    \begin{itemize}
        \item In global monitor topology, a single monitor is used to handle all the communication channels. This monitor randomly or systematically selects the communication channel and performs the PSL assertion-based verification. This topology exhibits the minimum area and power overhead but it decreases the HT detection accuracy.
        \item In channel-based topology, a monitor is attached to each communication channel. This topology exhibits the maximum HT detection accuracy but the area and power overheads are huge as compare to the global monitor. 
        \item In region-based topology, the SoC is divided into such small regions such that each region has the same number of communication channels. Then for each region a global monitor is inserted. This topology exhibits better HT detection accuracy as compared to global monitor, and better area and power overhead as compared to the channel-based topology.    
        
    \end{itemize}
    It is important to note that based on the criticality of the SoC regions, different topologies can be used for a different portion of the SoC. For example, the global monitors can be used in the regions which are not vulnerable to security attacks, and the channel-based topology can be used in the most vulnerable region.  
\end{enumerate}

\section{Comparative Discussion with state-of-the-art Techniques} \label{comparison} 

The run-time property checking technique can be used to detect the HTs at the hardware or software level. The fundamental issue with this technique is to extract the properties that provide maximum coverage. To address this issue, researchers have used the specification or functionality as property. For example, Ngo et al.~\cite{ngo2015hardware} extract the properties from the functionality and specification of the IC to design hardware property checkers (HPC). Similarly, Bloom et al.~\cite{bloom2009support} proposed to check the liveliness property at the OS level.  However, both of these techniques are applicable to functional Trojans and cannot detect the information leakage Trojans because they do not affect the functionality. Therefore, to detect such Trojans, we propose to exploit the statistical behavior of the communication for extracting the properties which can be used to detect functional and information leakage Trojans.

Table \ref{tab:tab4} provides a comparison of SIMCom with the state-of-the-art techniques that are relevant. The comparative analysis is based on five parameters. \textit{Overhead} provides the count of the extra component requirement for run-tim HT detection. \textit{Detection Approaches}, describes how a particular technique is detecting or protecting the intrusions that have any effect on SoC communication. \textit{Attributes}, indicates the key parameters used by a technique to protect or detect the intrusions. The final two comparison parameters describe the  \textit{Pros and Cons} of a particular technique. The proposed technique is found to be better than other state-of-the-art techniques in the following ways:
\begin{table}[!t]
	\centering
	\caption{Comparative discussion with the state-of-the-art Techniques}
	\label{tab:tab4}
	\resizebox{1\textwidth}{!}{
		\begin{tabular}{|c|c|c|c|l|l|}
			\hline
			\textbf{Techniques} & \begin{tabular}[c]{@{}c@{}}\textbf{Run-time}\\ \textbf{Overhead}\end{tabular} & \begin{tabular}[c]{@{}c@{}}\textbf{Detection}\\ \textbf{Approaches}\end{tabular} & \textbf{Attributes} & \multicolumn{1}{c|}{\textbf{Pros}} & \multicolumn{1}{c|}{\textbf{Cons}} \\ \hline
			\begin{tabular}[c]{@{}c@{}}\textbf{Kim et al.} \\ \cite{kim2011system} \end{tabular} & \begin{tabular}[c]{@{}c@{}} MUX based \\Wrappers,\\ Registers \end{tabular} & \begin{tabular}[c]{@{}c@{}}Protecting \\ Wrappers, \\ Hidden \\ Signals \end{tabular} & \begin{tabular}[c]{@{}c@{}}Master ID,\\ Restricted \\Addressing,\\ Ready and \\Grant Signals \end{tabular} & \begin{tabular}[c]{@{}l@{}}1. Low On-chip area \\ overhead \\ 2. Applicable to the SoC \\bus intrusions\end{tabular} & \begin{tabular}[c]{@{}l@{}}1. If the intrusion uses the \\ communication network\\ without affecting protocol,\\ then it fails. \end{tabular} \\ \hline
			
			\begin{tabular}[c]{@{}c@{}}\textbf{Kulkarni et al.} \\ \cite{kulkarni2016adaptive,kulkarni2016svm} \end{tabular} & \begin{tabular}[c]{@{}c@{}} Trained\\ ML Model \end{tabular} & \begin{tabular}[c]{@{}c@{}}ML Tools \\i.e., k-NN,\\ MBW, SVM\end{tabular} & \begin{tabular}[c]{@{}c@{}}Source and\\ Destination \\Core, Packet\\ Transfer Path,\\ Distance\end{tabular} & \begin{tabular}[c]{@{}l@{}}1. Low Latency \end{tabular} & \begin{tabular}[c]{@{}l@{}}1. Cannot detect Trojans\\ if the IPs are intruded.\\ 2. Only applicable if \\intrusion affects the \\communication protocols \end{tabular} \\ \hline
			
			\begin{tabular}[c]{@{}c@{}}\textbf{Ngo et al.} \\ \cite{ngo2015hardware} \end{tabular} & \begin{tabular}[c]{@{}c@{}} Assertions \end{tabular} & \begin{tabular}[c]{@{}c@{}}PSL\\ Assertions\end{tabular} & \begin{tabular}[c]{@{}c@{}}Functional \\Specifications\end{tabular} & \begin{tabular}[c]{@{}l@{}}1. Low area and power\\ overhead \end{tabular} & \begin{tabular}[c]{@{}l@{}}1. Cannot detect Trojans\\ that steals the information.\end{tabular} \\ \hline
			
			\begin{tabular}[c]{@{}c@{}}\textbf{ SIMCom} \end{tabular} & \begin{tabular}[c]{@{}c@{}} $H$, $P$ and $\sigma$\\ computing\\ blocks, \\ Four 3.2KB \\ Memories  \end{tabular} & \begin{tabular}[c]{@{}c@{}}PSL \\Assertions\end{tabular} & \begin{tabular}[c]{@{}c@{}}Hurst exponent,\\ Hop Distribution,\\ Standard Deviation \\of Injection \\ Distribution\end{tabular} & \begin{tabular}[c]{@{}l@{}}1. Detects the Trojans in \\3PIP modules that use \\SoC communication\\ network \\  2. It also detects Trojans,\\ which do not affect  the \\protocol directly.   \end{tabular} & \begin{tabular}[c]{@{}l@{}}1. Only applicable if\\ intrusion affects the \\SoC communication\end{tabular} \\ \hline			
	\end{tabular}}
\end{table}
\begin{enumerate}
	\item The proposed technique does not require any wrappers and hidden control signals, like the LOCK signal, for protecting against the Trojans, as some other techniques require, e.g., \cite{kim2011system}.
	
	\item Unlike \cite{kulkarni2016adaptive,kulkarni2016svm}, our technique does not presume that Trojan payload activation provides a very high change in the communication channel or any change in communication protocols. Moreover, it can also detect the Trojans, which have indirect effects on communication channels.
	
	\item The proposed technique can be used for any known bus system by adding the $H$, $P$, and $\sigma$ measurement blocks and PSL assertions obtained from the communication behavior of the pre-market test stage. 
	
	\item SIMCom is a hardware property checking (HPC) based HT detection technique at the hardware-level like proposed by Ngo et al.~\cite{ngo2015hardware}. However, unlike the existing HPC-based technique~\cite{ngo2015hardware} is based on the functional properties of the protocol. Therefore, it is unable to detect the Trojans which leak the information without affecting the communication protocols. On the other hand, SIMCom detects the functional Trojans as well as the information leakage Trojans. Figures~\ref{fig:FP_FN_8051} and~\ref{fig:FP_FN_LEON3} show that HT detection accuracy of the existing HPC-based technique~\cite{ngo2015hardware} for most of the AES Trojans is 0 because these Trojans leak the information without changing the functionality. The detailed hardware implementation of SIMCom is given in Figure~\ref{fig:HW}.
\end{enumerate}

\section{Conclusions}\label{conclusion}
This paper utilizes a statistical traffic modeling of SoC for sniffing HTs during run-time. The proposed methodology consists of two major steps: The first step is to extract the pre-market test stage traffic model by computing three parameters, i.e., Hurst exponent ($H$), spatial hop distribution ($P$) and spatial injection distribution ($\sigma$). Then, this extracted behavior is translated to its corresponding PSL assertions, which are embedded into the given SoC along with the blocks, which compute the traffic modeling parameters. For illustration purpose, for three SoCs, i.e., SoC1 ( four single-core MC8051 and UART modules), SoC2 (four single-core MC8051, AES, ethernet, memctrl, BasicRSA, RS232 modules), and SoC3 (four single-core LEON3 connected with each other and AES, ethernet, memctrl, BasicRSA, RS23s modules microcontrollers). For validation purposes, we used all available benchmarks on Trust-Hub. The experimental results show that the proposed methodology is able to detect all intrusions with less than 1\% area and power overhead.  To the best of our knowledge, this is the first time that a statistical sniffing of SoC's communication traffic based approach is utilized to detect hardware Trojans in 3PIP modules.

\section*{Acknowledgment}
This work is supported in parts by the Austrian Research Promotion Agency (FFG) and the Austrian Federal Ministry for Transport, Innovation, and Technology (BMVIT) under the “ICT of the Future” project, IoT4CPS: Trustworthy IoT for Cyber-Physical Systems.

\bibliographystyle{elsarticle-num}
\bibliography{SIMCom.bbl}

\begin{thebibliography}{10}
\expandafter\ifx\csname url\endcsname\relax
  \def\url#1{\texttt{#1}}\fi
\expandafter\ifx\csname urlprefix\endcsname\relax\def\urlprefix{URL }\fi
\expandafter\ifx\csname href\endcsname\relax
  \def\href#1#2{#2} \def\path#1{#1}\fi

\bibitem{mohan2013s3a}
S.~Mohan, S.~Bak, E.~Betti, H.~Yun, L.~Sha, M.~Caccamo, S3a: Secure system
  simplex architecture for enhanced security and robustness of cyber-physical
  systems, in: Proceedings of the 2nd ACM international conference on High
  confidence networked systems, ACM, 2013, pp. 65--74.

\bibitem{levshun2019design}
D.~Levshun, A.~Chechulin, I.~Kotenko, Y.~Chevalier, Design and verification
  methodology for secure and distributed cyber-physical systems, in: 2019 10th
  IFIP International Conference on New Technologies, Mobility and Security
  (NTMS), IEEE, 2019, pp. 1--5.

\bibitem{wang2018model}
P.~Wang, J.~Liu, J.~Lin, C.-H. Chu, Model based energy consumption analysis of
  wireless cyber physical systems, Journal of Signal Processing Systems
  90~(8-9) (2018) 1191--1204.

\bibitem{giakoumis2018chaos}
A.~Giakoumis, C.~K. Volos, I.~N. Stouboulos, H.~Polatoglou, I.~M. Kyprianidis,
  Chaos generator device based on a 32 bit microcontroller embedded system, in:
  2018 7th International Conference on Modern Circuits and Systems Technologies
  (MOCAST), IEEE, 2018, pp. 1--4.

\bibitem{ratasich2019roadmap}
D.~Ratasich, F.~Khalid, F.~Geissler, R.~Grosu, M.~Shafique, E.~Bartocci, A
  roadmap toward the resilient internet of things for cyber-physical systems,
  IEEE Access 7 (2019) 13260--13283.

\bibitem{shafique2018intelligent}
M.~Shafique, F.~Khalid, S.~Rehman, Intelligent security measures for smart
  cyber physical systems, in: 2018 21st Euromicro Conference on Digital System
  Design (DSD), IEEE, 2018, pp. 280--287.

\bibitem{moura2019cyber}
J.~Moura, D.~Hutchison, Cyber-physical systems resilience: State of the art,
  research issues and future trends, arXiv preprint arXiv:1908.05077.

\bibitem{tehranipoor2010survey}
M.~Tehranipoor, A {S}urvey of {H}ardware {T}rojan {T}axonomy and {D}etection,
  Design \& Test Computer 27~(1) (2010) 10--25.

\bibitem{subramani2018hardware}
K.~S. Subramani, A.~Antonopoulos, A.~Nosratinia, Y.~Makris, Hardware trojans in
  wireless networks, Ph.D. thesis, University of Texas at Dallas (2018).

\bibitem{villasenor2013hidden}
J.~Villasenor, M.~Tehranipoor, The hidden dangers of chop-shop electronics:
  Clever counterfeiters sell old components as new threatening both military
  and commercial systems, IEEE Spectrum (cover story) 2013.

\bibitem{bloomberg_security}
J.~Robertson, M.~Riley, {The Big Hack: How China Used a Tiny Chip to Infiltrate
  U.S. Companies} (2018).

\bibitem{zarrinchian2017latch}
G.~Zarrinchian~et.al., Latch-based structure: A high resolution and
  self-reference technique for hardware trojan detection, Trans. C 66~(1)
  (2017) 100--113.

\bibitem{plusquellic2018detecting}
J.~Plusquellic, et~al., Detecting hardware trojans using delay analysis, in:
  The Hardware Trojan War, Springer, 2018, pp. 219--267.

\bibitem{nandhini2018delay}
S.~K. Nandhini, S.~Vallinayagam, H.~Harshitha, V.~C.~S. Azad, N.~Mohankumar,
  Delay-based reference free hardware trojan detection using virtual
  intelligence, in: Information Systems Design and Intelligent Applications,
  Springer, 2018, pp. 506--514.

\bibitem{fang2018prefetch}
H.~Fang, S.~S. Dayapule, F.~Yao, M.~Doroslova{\v{c}}ki, G.~Venkataramani,
  Prefetch-guard: Leveraging hardware prefetches to defend against cache timing
  channels, in: 2018 IEEE International Symposium on Hardware Oriented Security
  and Trust (HOST), IEEE, 2018, pp. 187--190.

\bibitem{xue2018hardware}
H.~Xue, S.~Ren, Hardware trojan detection by timing measurement: Theory and
  implementation, Microelectronics journal 77 (2018) 16--25.

\bibitem{lodhi2017power}
F.~K. Lodhi, S.~R. Hasan, O.~Hasan, F.~Awwadl, Power profiling of
  microcontroller's instruction set for runtime hardware trojans detection
  without golden circuit models, in: Design, Automation \& Test in Europe
  Conference \& Exhibition (DATE), 2017, IEEE, 2017, pp. 294--297.

\bibitem{hoque2017golden}
T.~Hoque, S.~Narasimhan, X.~Wang, S.~Mal-Sarkar, S.~Bhunia, Golden-free
  hardware trojan detection with high sensitivity under process noise, Journal
  of Electronic Testing 33~(1) (2017) 107--124.

\bibitem{lodhi2016self}
F.~K. Lodhi, I.~Abbasi, F.~Khalid, O.~Hasan, F.~Awwad, S.~R. Hasan, A
  self-learning framework to detect the intruded integrated circuits, in: 2016
  IEEE International Symposium on Circuits and Systems (ISCAS), IEEE, 2016, pp.
  1702--1705.

\bibitem{zhang2018data}
Y.~Zhang, H.-d. Quan, X.-w. Li, Data activated processor hardware trojan
  detection using differential bit power analysis, in: Proceedings of the 2nd
  International Conference on Cryptography, Security and Privacy, ACM, 2018,
  pp. 134--137.

\bibitem{gbade2014signature}
A.~Gbade-Alabi, D.~Keezer, V.~Mooney, A.~Y. Poschmann, M.~St{\"o}ttinger,
  K.~Divekar, A {S}ignature {B}ased {A}rchitecture for {T}rojan {D}etection,
  in: Embedded Systems Security, 2014, p.~3.

\bibitem{lodhi2014hardware}
F.~K. Lodhi, S.~R. Hasan, O.~Hasan, F.~Awwad, Hardware {T}rojan {D}etection in
  {S}oft {E}rror {T}olerant {M}acro {S}ynchronous {M}icro {A}synchronous
  ({MSMA}) pipeline, in: Midwest Symposium on Circuits and Systems, 2014, pp.
  659--662.

\bibitem{zhang2014detrust}
J.~Zhang, F.~Yuan, Q.~Xu, Detrust: {D}efeating {H}ardware {T}rust
  {V}erification with {S}tealthy {I}mplicitly-{T}riggered {H}ardware {T}rojans,
  in: Conference on Computer \& Communications Security, 2014, pp. 153--166.

\bibitem{waksman2013fanci}
A.~Waksman, M.~Suozzo, S.~Sethumadhavan, Fanci: {I}dentification of {S}tealthy
  {M}alicious {L}ogic using {B}oolean {F}unctional {A}nalysis, in: Computer \&
  communications security, 2013, pp. 697--708.

\bibitem{zhang2015veritrust}
J.~Zhang, F.~Yuan, L.~Wei, Y.~Liu, Q.~Xu, Transactions on Computer-Aided Design
  of Integrated Circuits \& Systems 34~(7) (2015) 1148--1161.

\bibitem{haider2014hatch}
S.~K. Haider, C.~Jin, M.~Ahmad, D.~M. Shila, O.~Khan, M.~V.~Dijk, Hatch:
  {H}ardware {T}rojan {C}atcher, IACR Cryptology ePrint 2014 (2014) 943.

\bibitem{ngo2015hardware}
X.-T. Ngo, et~al., Hardware trojan detection by delay and electromagnetic
  measurements, in: Proceedings of the 2015 Design, Automation \& Test in
  Europe Conference \& Exhibition, EDA Consortium, 2015, pp. 782--787.

\bibitem{zareen2018detecting}
F.~Zareen, R.~Karam, Detecting rtl trojans using artificial immune systems and
  high level behavior classification, in: 2018 Asian Hardware Oriented Security
  and Trust Symposium (AsianHOST), IEEE, 2018, pp. 68--73.

\bibitem{cui2018hardware}
X.~Cui, E.~Koopahi, K.~Wu, R.~Karri, Hardware trojan detection using the order
  of path delay, ACM Journal on Emerging Technologies in Computing Systems
  (JETC) 14~(3) (2018) 33.

\bibitem{liu2019hardware}
Y.~Liu, J.~He, H.~Ma, Y.~Zhao, Hardware trojan detection leveraging a novel
  golden layout model towards practical applications, Journal of Electronic
  Testing (2019) 1--13.

\bibitem{he2017hardware}
J.~He, Y.~Zhao, X.~Guo, Y.~Jin, Hardware trojan detection through chip-free
  electromagnetic side-channel statistical analysis, IEEE Transactions on Very
  Large Scale Integration (VLSI) Systems 25~(10) (2017) 2939--2948.

\bibitem{pelgrom2017nyquist}
M.~Pelgrom, Nyquist analog-to-digital conversion, in: Analog-to-Digital
  Conversion, Springer, 2017, pp. 285--403.

\bibitem{bhunia2014hardware}
S.~Bhunia, M.~S. Hsiao, M.~Banga, S.~Narasimhan, Hardware {T}rojan {A}ttacks:
  {T}hreat {A}nalysis and {C}ountermeasures, Proceedings of the IEEE 102~(8)
  (2014) 1229--1247.

\bibitem{becker2013stealthy}
G.~T. Becker, F.~Regazzoni, C.~Paar, W.~P. Burleson, Stealthy dopant-level
  hardware trojans, in: International Workshop on Cryptographic Hardware and
  Embedded Systems, Springer, 2013, pp. 197--214.

\bibitem{hasan2015tenacious}
S.~R. Hasan, S.~F. Mossa, O.~S.~A. Elkeelany, F.~Awwad, Tenacious {H}ardware
  {T}rojans due to {H}igh {T}emperature in {M}iddle {T}iers of 3-{D} {IC}s, in:
  Midwest Symposium on Circuits \& Systems, 2015, pp. 1--4.

\bibitem{mossa2017hardware}
S.~F. Mossa, S.~R. Hasan, O.~Elkeelany, Hardware trojans in 3-d ics due to nbti
  effects and countermeasure, Integration 59 (2017) 64--74.

\bibitem{mossa2017self}
S.~F. Mossa, S.~R. Hasan, O.~Elkeelany, Self-triggering hardware trojan: Due to
  nbti related aging in 3-d ics, Integration 58 (2017) 116--124.

\bibitem{forte2013temperature}
D.~Forte, C.~Bao, A.~Srivastava, Temperature {T}racking: {An} {I}nnovative
  {R}un-time {A}pproach for {H}ardware {T}rojan {D}etection, in: Computer-Aided
  Design, 2013, pp. 532--539.

\bibitem{bao2015temperature}
C.~Bao, D.~Forte, A.~Srivastava, Temperature{T}racking: {T}oward {R}obust
  {R}un-time {D}etection of {H}ardware {T}rojans, Transactions on
  Computer-Aided Design of Integrated Circuits \& Systems 34~(10) (2015)
  1577--1585.

\bibitem{zhao2015applying}
H.~Zhao, K.~Kwiat, C.~Kamhoua, M.~Rodriguez, Applying {C}haos {T}heory for
  {R}untime {H}ardware {T}rojan {D}etection, in: Computational Intelligence for
  Security \& Defense Applications, 2015, pp. 1--6.

\bibitem{bao2016reverse}
C.~Bao, D.~Forte, A.~Srivastava, On {R}everse {E}ngineering-based {H}ardware
  {T}rojan {D}etection, Transactions on Computer-Aided Design of Integrated
  Circuits \& Systems 35~(1) (2016) 49--57.

\bibitem{iwase2015detection}
T.~Iwase, Y.~Nozaki, M.~Yoshikawa, T.~Kumaki, Detection {T}echnique for
  {H}ardware {T}rojans using {M}achine {L}earning in {F}requency {D}omain, in:
  Conference on Consumer Electronics, 2015, pp. 185--186.

\bibitem{trust-HUB}
M.~Tehranipoor, H.~Salamani, \href{https://www.trust-hub.org/}{{trust-HUB}}
  (2016).
\newline\urlprefix\url{https://www.trust-hub.org/}

\bibitem{soteriou2006statistical}
V.~Soteriou, H.~Wang, L.~Peh, A {S}tatistical {T}raffic {M}odel for {O}n-{C}hip
  {I}nterconnection {N}etworks, in: Symposium on Modeling, Analysis, \&
  Simulation, 2006, pp. 104--116.

\bibitem{kleinow2002testing}
H.~D. M.~T. Kleinow, Testing {C}ontinuous {T}ime {M}odels in {F}inancial
  {M}arkets, Ph.D. thesis, Humboldt-Universit{\"a}t zu Berlin (2002).

\bibitem{xiao2016hardware}
K.~Xiao, D.~Forte, Y.~Jin, R.~Karri, S.~Bhunia, M.~Tehranipoor, Hardware
  {T}rojans: {L}essons {L}earned after {O}ne {D}ecade of {R}esearch,
  Transactions on Design Automation of Electronic Systems 22~(1) (2016) 6.

\bibitem{cui2014high}
X.~Cui, K.~Ma, L.~Shi, K.~Wu, High-level synthesis for run-time hardware trojan
  detection and recovery, in: 2014 51st ACM/EDAC/IEEE Design Automation
  Conference (DAC), IEEE, 2014, pp. 1--6.

\bibitem{sutikno2012simplified}
T.~Sutikno, A.~Z. Jidin, A.~Jidin, N.~R.~N. Idris, Simplified vhdl coding of
  modified non-restoring square root calculator, International Journal of
  Reconfigurable and Embedded Systems 1~(1) (2012) 37.

\bibitem{kreutz2001communication}
M.~E. Kreutz, L.~Carro, C.~A. Zeferino, A.~A. Susin, Communication
  {A}rchitectures for {S}ystem-on-{C}hip, in: Integrated Circuits \& Systems
  Design, 2001, pp. 14--19.

\bibitem{wang2008detecting}
X.~Wang, M.~Tehranipoor, J.~Plusquellic, Detecting malicious inclusions in
  secure hardware: Challenges and solutions, in: 2008 IEEE International
  Workshop on Hardware-Oriented Security and Trust, IEEE, 2008, pp. 15--19.

\bibitem{brasser2018special}
F.~Brasser, L.~Davi, A.~Dhavlle, T.~Frassetto, S.~M.~P. Dinakarrao,
  S.~Rafatirad, A.-R. Sadeghi, A.~Sasan, H.~Sayadi, S.~Zeitouni, et~al.,
  Special session: Advances and throwbacks in hardware-assisted security, in:
  2018 International Conference on Compilers, Architectures and Synthesis for
  Embedded Systems (CASES), IEEE, 2018, pp. 1--10.

\bibitem{gil2012studying}
D.~Gil-Tom{\'a}s, J.~Gracia-Mor{\'a}n, J.-C. Baraza-Calvo, L.-J. Saiz-Adalid,
  P.-J. Gil-Vicente, Studying the effects of intermittent faults on a
  microcontroller, Microelectronics Reliability 52~(11) (2012) 2837--2846.

\bibitem{korak2014effects}
T.~Korak, M.~Hoefler, On the effects of clock and power supply tampering on two
  microcontroller platforms, in: 2014 Workshop on Fault Diagnosis and Tolerance
  in Cryptography, IEEE, 2014, pp. 8--17.

\bibitem{khalid2018runtime}
F.~Khalid, S.~R. Hasan, O.~Hasan, F.~Awwad, Runtime hardware trojan monitors
  through modeling burst mode communication using formal verification,
  Integration 61 (2018) 62--76.

\bibitem{bloom2009support}
G.~Bloom, B.~Narahari, R.~Simha, Os support for detecting trojan circuit
  attacks, in: 2009 IEEE International Workshop on Hardware-Oriented Security
  and Trust, IEEE, 2009, pp. 100--103.

\bibitem{kim2011system}
L.~Kim, J.~D. Villasenor, A {S}ystem-on-{C}hip {B}us {A}rchitecture for
  {T}hwarting {I}ntegrated {C}ircuit {T}rojan {H}orses, Transactions on Very
  Large Scale Integration Systems 19~(10) (2011) 1921--1926.

\bibitem{kulkarni2016adaptive}
A.~Kulkarni, Y.~Pino, T.~Mohsenin, Adaptive {R}eal-time {T}rojan {D}etection
  {F}ramework {T}hrough {M}achine {L}earning, in: Hardware Oriented Security \&
  Trust, 2016, pp. 120--123.

\bibitem{kulkarni2016svm}
A.~Kulkarni, Y.~Pino, T.~Mohsenin, {SVM}-based {R}eal-time {H}ardware {T}rojan
  {D}etection for {M}any-{C}ore {P}latform, in: Symposium on Quality Electronic
  Design, 2016, pp. 362--367.

\end{thebibliography}

\end{document}